\newcommand*{\ATLASLATEXPATH}{latex/}

\documentclass[UKenglish,texlive=2011,cernpreprint]{\ATLASLATEXPATH atlasdoc}

\usepackage{\ATLASLATEXPATH atlaspackage}

\usepackage{\ATLASLATEXPATH atlasbiblatex}

\usepackage[hion=true]{\ATLASLATEXPATH atlasphysics}

\graphicspath{{logos/}{figures/}}

\usepackage{atlas_tables-defs}
\usepackage{RpPb_paper_2012-defs}

\AtlasTitle{Transverse momentum, rapidity, and centrality dependence of inclusive charged-particle production in \sqn=5.02\,\TeV\ \pPb collisions measured by the ATLAS experiment}

\author{The ATLAS Collaboration}

\AtlasJournal{\PLB}
\PreprintIdNumber{CERN-EP-2016-077}
\AtlasDate{\today}
\AtlasJournalRef{\PLB\ 763 (2016) 313-336}
\AtlasDOI{10.1016/j.physletb.2016.10.053}
\arXivId{1605.06436}

\AtlasAbstract{%
  
Measurements of the per-event charged-particle yield as a function of the charged-particle transverse momentum and rapidity are performed using \pPb collision data collected by the ATLAS experiment at the LHC at a centre-of-mass energy of $\sqn=5.02$\,\TeV. Charged particles are reconstructed over pseudorapidity $|\eta|<2.3$ and transverse momentum between $0.1$~\GeV\ and $22$~\GeV\ in a dataset corresponding to an integrated luminosity of $1$~$\mu {\textrm b}^{-1}$. The results are presented in the form of charged-particle nuclear modification factors, where the \pPb charged-particle multiplicities are compared between central and peripheral \pPb collisions as well as to charged-particle cross sections measured in \pp collisions. The \pPb collision centrality is characterized by the total transverse energy measured in  ${-4.9<\eta<-3.1}$, which is in the direction of the outgoing lead beam. Three different estimations of the number of nucleons participating in the \pPb collision are carried out using the Glauber model and two Glauber--Gribov colour-fluctuation extensions to the Glauber model. The values of the nuclear modification factors are found to vary significantly as a function of rapidity and transverse momentum. A broad peak is observed for all centralities and rapidities in the nuclear modification factors for charged-particle transverse momentum values around $3$~\GeV. The magnitude of the peak increases for more central collisions as well as rapidity ranges closer to the direction of the outgoing lead nucleus.

}

\hypersetup{pdftitle={ATLAS document},pdfauthor={The ATLAS Collaboration}}

\begin{document}

\maketitle

\tableofcontents

\clearpage

\section{Introduction} 
\label{sec:intro}

Proton--nucleus collisions at ultrarelativistic energies provide an opportunity to understand the role of the nuclear environment in modifying hard scattering rates. 
Several physics effects are expected to induce deviations from a simple proportionality between the scattering rate and the number of binary
nucleon--nucleon collisions~\cite{Albacete:2013ei}. 
First, nuclear shadowing effects have long been observed in deep-inelastic scattering on nuclei, as well as in proton-nucleus collisions,
indicating that nucleons embedded in a nucleus have a modified structure. This modification tends to suppress
hadron production at low to moderate momentum, and is addressed by a variety of theoretical approaches~\cite{Kopeliovich:1999am,Qiu:2004da}. 
Some of these approaches describe hadron production cross sections in terms of a universal set of nuclear parton distribution functions (nPDF), which are parameterized as modifications to the free nucleon PDFs~\cite{Eskola:2009uj, Eskola:1998iy,Helenius:2012wd,Hirai:2007sx,deFlorian:2011fp,McLerran:1993ni, Kharzeev:2003wz, Albacete:2010bs, Tribedy:2011aa}. 
Second, energy loss in ``cold nuclear matter'' is expected to modify hadron production rates at high transverse
momentum (\pT)~\cite{Kang:2012kc,JalilianMarian:2011dt,Albacete:2012xq,Rezaeian:2012ye}. 
Third, a relative enhancement of hadron production rates at moderate momenta is observed in proton--nucleus collisions~\cite{PhysRevD.11.3105}, 
which can be attributed to initial-state scattering of the incoming nucleon~\cite{Kopeliovich:2002yh,Vitev:2003xu} or radial flow effects~\cite{Werner:2013tya}. 
Finally, the appearance of “ridge-like” structures in high-multiplicity \pp and \pPb events~\cite{Khachatryan:2010gv,CMS:2012qk,Abelev:2012ola,Abelev:2012ola,Aad:2013fja,Chatrchyan:2013nka} suggests that small collision systems have the same hydrodynamic origin as \PbPb events~\cite{Bozek:2013uha}, or that there are already strong correlations in the initial state from gluon saturation~\cite{Dusling:2013qoz}. All of these effects can be explored experimentally by the measurement of charged-hadron production as a function of transverse momentum.

For proton--lead (\pPb) collisions, assuming that the initial parton densities are the incoherent superposition of the nucleonic parton densities, the per-event particle production yield can be estimated by the product $\sigma_{_\mathrm{NN}}\times\avgTpb$. Here $\sigma_{_\mathrm{NN}}$ is the cross section for the analogous nucleon--nucleon collision process and \avgTpb is the average value of the nuclear thickness function over a distribution of the impact parameters of protons incident on the nuclear target. It can be thought of as a per-collision luminosity.
The nuclear modification factor, \rpPb, is defined as the ratio of the measured charged-particle production yield
in \pPb collisions, normalized by \avgTpb, to the cross section of the charged-particle production yield in \pp collisions: 
\begin{equation}
\rpPb(\pT,\ystar) = \frac{1}{\avgTpb}\frac{1/\Nevt~\mathrm{d}^{2}N_{ p\text{Pb}} / \mathrm{d}\ystar \mathrm{d}\pT}{\mathrm{d}^2\mathrm{\sigma}_{\pp} / \mathrm{d}\ystar \mathrm{d}\pT},
  \label{eq:rpA}
\end{equation}
where $N_{\mathrm{evt}}$ is the number of \pPb events, $\mathrm{d}^{2}N_{ p\text{Pb}} / \mathrm{d}\ystar \mathrm{d}\pT$ is the differential yield of charged particles in \pPb collisions and $\mathrm{d}^2\mathrm{\sigma}_{\pp} / \mathrm{d}\ystar \mathrm{d}\pT$ is the differential charged-particle production cross section in \pp collisions. Both numerator and denominator are presented in terms of \ystar, the rapidity in the nucleon--nucleon centre-of-mass frame. 
In the absence of initial-state and nuclear effects, the ratio \rpPb is expected to be unity at high \pT~\cite{Barnafoldi:2011px}. 
Another measure of nuclear modification is the quantity \rcp, which is defined to be:
\begin{equation}
\rcp(\pT,\eta) = \frac{\langle T_{\mathrm{Pb,P}}\rangle}{\langle T_{\mathrm{Pb},\mathrm{C}}\rangle} \frac{(1/N_{\mathrm{evt,C}}) \mathrm{d}^2 N_{{p\text{Pb},\mathrm{C}}} / \mathrm{d}\eta \mathrm{d}\pT}{ (1/N_{\mathrm{evt,\mathrm{P}}}) \mathrm{d}^2 N_{{p\text{Pb},\mathrm{P}}} / \mathrm{d}\eta \mathrm{d}\pT},
\label{eq:rcp}
\end{equation}
and can be constructed without the need for a \pp reference spectrum. The indices ``P'' and ``C'' label peripheral (large impact parameter) and central (small impact parameter) centrality intervals, respectively. The \rcp is presented as a function of pseudorapidity ($\eta$) rather than $\ystar$ since both numerator and denominator are from the same colliding systems. 
Measurements of \rpPb and \rcp provide useful input for constraining models of shadowing, energy loss and radial flow effects. They should also provide useful input for the determination of nuclear parton distribution functions, in particular as a function of proton impact parameter~\cite{Helenius:2012wd}. The absolute values of the nuclear modification depend on the \avgTpb values and should be interpreted with respect to the assumptions underlying the particular model used to calculate the normalisation.

A recent ATLAS publication~\cite{Aad:2015zza} has reported measurements of the mean charged-particle multiplicity as a function of pseudorapidity and collision centrality and explored the relationship between the centrality dependence of the particle production and models of the initial nuclear geometry. The results presented here utilize the same centrality definition and geometric models, but build upon that work by exploring the \pT, \eta and \ystar dependence of per-event charged-particle yields in \pPb collisions  at a centre-of-mass energy $\sqn=5.02$\,\TeV\ and comparing that dependence to the expectations from \pp collisions through the quantities \rpPb and \rcp.

These measurements are an extension of a similar programme carried out 
at the Relativistic Heavy Ion Collider, where all experiments reported the absence of charged-particle 
suppression at $2<\pT<10$\,\GeV\ in \dAu collisions~\cite{Arsene:2003yk,PhysRevLett.91.072303,Back:2003ns,PhysRevLett.91.072304,Adams:2006nd,Adare:2013esx}, in contrast to the strong suppression found in \AuAu collisions \cite{PhysRevLett.91.072304,PhysRevLett.91.072303}. Measurements of nuclear modification factors as a function of transverse momentum in a narrow pseudorapidity window relative to the centre-of-mass frame $|\eta_{\mathrm{CM}}|<0.3$ have been reported by ALICE integrated over centrality~\cite{ALICE:2012mj,Abelev:2014dsa} and differentially for several centrality classes~\cite{Adam:2014qja,Adam:2016dau}. Similarly, CMS results have been reported integrated over centrality and in a broader pseudorapidity window, $|\eta_{\mathrm{CM}}|<1$~\cite{CMS-PAS-HIN-12-017}.

\section{The ATLAS detector} 
\label{sec:exper}

The ATLAS detector~\cite{Aad:2008zzm} at the Large Hadron Collider (LHC) covers almost the entire solid angle\footnote{\AtlasCoordFootnote} around the collision point. It consists of an inner tracking detector surrounded by a thin superconducting solenoid, electromagnetic and hadronic calorimeters, and a muon spectrometer incorporating three large superconducting toroidal magnets.

The inner detector (ID) system is immersed in a \SI{2}{\tesla} axial magnetic field 
and provides charged-particle tracking in the pseudorapidity range $|\eta|<2.5$. 
The ID tracker is composed of three detector subsystems. 
Closest to the interaction point is a high-granularity silicon pixel detector 
covering $|\eta|<2.7$, which typically provides three measurements per track. 
Next is a silicon microstrip tracker (SCT), which typically yields four pairs of hits per track, each providing a two-dimensional measurement point. 
The silicon detectors are complemented by the straw-tube transition radiation tracker,
which enables radially extended track reconstruction up to $|\eta| = 2.0$. 

The calorimeter system covers the pseudorapidity range $|\eta|<4.9$.
Within the region $|\eta|< 3.2$, electromagnetic calorimetry is provided by 
high-granularity lead/liquid-argon (LAr) electromagnetic calorimeters,
with an additional thin LAr presampler covering $|\eta|<1.8$,
to measure the contribution of showers initiated in the material upstream of the calorimeters.
Hadronic calorimetry is provided by a steel/scintillator-tile calorimeter,
segmented into three barrel structures within $|\eta|<1.7$, and two copper/LAr hadronic endcap calorimeters covering $1.5<|\eta|<3.2$.
The calorimeter coverage is completed with forward copper/LAr and tungsten/LAr calorimeter modules
optimized for electromagnetic and hadronic measurements, respectively, covering $3.1<|\eta|<4.9$. 
The minimum-bias trigger scintillators (MBTS) detect charged particles over $2.1<|\eta|<3.9$ using two hodoscopes, each of which is subdivided into 16 counters positioned at $z=\pm3.6$~m. 

A three-level trigger system is used to select events~\cite{PERF-2011-02}.
The Level-1 trigger is implemented in hardware and uses a subset of detector information
to reduce the event rate to $100$~kHz. This is followed by two software-based trigger levels which together reduce the event rate to about $1000$~Hz, which is recorded for data analysis.

\section{Datasets and event selection}
\label{sec:events}
\subsection{Event selection in $\pPb$ collisions}

The \pPb collisions were recorded by the ATLAS detector in September 2012 using a trigger that selected events with at least one hit in each side of the MBTS, with the resulting dataset corresponding to an integrated luminosity of $1$~$\mu {\textrm b}^{-1}$. During that run the LHC was configured with a clockwise $4$~\TeV\ proton beam and an anti-clockwise $1.57$~\TeV\ per-nucleon $^{208}$Pb beam that together produced collisions with a nucleon--nucleon centre-of-mass energy of $\sqs=5.02$~\TeV\ and a longitudinal rapidity boost of $\ylab=0.465$ units with respect to the ATLAS laboratory frame. Following a common convention used for \pA\ measurements, the rapidity is taken to be positive in the direction of the proton beam, i.e. opposite to the usual ATLAS convention for \pp\ collisions. With this convention, the ATLAS laboratory frame rapidity, $y$, and the \pPb centre-of-mass system rapidity, \ystar, are related by $\ystar = y-0.465$.

Charged-particle tracks and collision
vertices are reconstructed from clusters in the pixel detector 
and the SCT using an algorithm optimized for minimum-bias \pp\
measurements~\cite{Aad:2010ac}. 
The \pPb events are required to have a collision vertex satisfying
$|\zvtx|< 150$~mm, at least one hit in each side of the MBTS, and a
difference between the time measurements in the two MBTS hodoscopes of less than
10\,ns. Events containing multiple \pPb collisions (pile-up) are suppressed by 
rejecting events that contain a second reconstructed vertex with 
a scalar transverse momentum sum of associated tracks of $\Sigma\pT^{2}>5$ GeV.
The residual contamination from pile-up events has been estimated to be $10^{-4}$~\cite{Aad:2013fja}.

To remove contributions from electromagnetic 
and diffractive processes, a rapidity gap criterion is applied to the \pPb\ data using the procedure outlined in Ref.~\cite{Aad:2015zza}. The procedure utilizes energy deposits in the calorimeter identified using so-called topological clusters~\cite{Aad:2010ad}. The detector is divided into slices of $\Delta\eta=0.2$, and
``edge'' gaps are calculated as the distance from the edge of the
calorimeter ($\eta=-4.9$) to the nearest slice that contains a cluster with a
minimum transverse energy of 200 MeV. 
Events with a large edge gap ($\ygappb > 2$) in the negative $\eta$ (Pb) direction are excluded from the analysis. 
The gap requirement removes $1$\% of the events passing the vertex and MBTS 
timing cuts, which yields a total of $2.1\times10^{6}$ events used for further analysis.

\subsection{Event selection in \pp collisions}
The \pp\ spectrum used as a reference for the \pPb measurement is based on an 
interpolation of two data samples taken at $\sqs=2.76$~\TeV\ and 7~\TeV. 
Proton--proton collisions at $\sqs = 2.76$~\TeV\ with total integrated luminosity 200~nb$^{-1}$ were obtained 
by the ATLAS experiment in March 2011. Proton--proton collisions at $\sqs = 7$~\TeV\ with total integrated luminosity  130~$\mu$b$^{-1}$ were obtained in April 2010. In both cases, the trigger selected events with at least one hit in the MBTS detector. 
The average number of collisions per bunch crossing during these data-taking periods is 0.4 and 0.01 for the $\sqs=2.76$~\TeV\ and $\sqs=7$~\TeV\ datasets, respectively. 
Events are required to satisfy the same $\zvtx$ and MBTS requirements as for \pPb analysis. 
\subsection{Monte Carlo event simulation}
\label{sec:mcset}
The response of the ATLAS detector and the performance of
reconstruction algorithms are evaluated using one million simulated minimum-bias
\pPb events at $\sqs=5.02$~\TeV, produced by version 1.38b of the \HIJING event
generator \cite{Wang:1991hta}. Diffractive processes are disabled. 
To match the LHC \pPb beam conditions, the four-momentum 
of each generated particle is longitudinally boosted by a rapidity of $-0.465$. The generator-level events are then passed through a \GEANT4 simulation of the ATLAS detector~\cite{Agostinelli:2002hh,Aad:2010ah}. The simulated events are digitized using data conditions appropriate to the \pPb run and are reconstructed using the same algorithms that are applied to the experimental data.

For the \pp\ analysis, $20$ million events were produced using the $\PYTHIA6$~\cite{Sjostrand:2006za} event generator with the AUET2B parameter set~\cite{ATL-PHYS-PUB-2011-009} at both $\sqs=2.76$~\TeV\ and $7$~\TeV\ (with versions $6.423$ and $6.421$ respectively). Additional samples produced using $\PYTHIA8$~\cite{Sjostrand:2007gs} with the 4C parameter set~\cite{Katzy2013141}, and \HERWIGpp with the UEEE5 parameter set~\cite{Bahr:2008pv}, are used for studying systematic uncertainties (see Sections~\ref{sec:ppref} and~\ref{sec:syst}). 

\section{Centrality selection}
\label{sec:centrality}
The centrality determination for \pPb collisions in ATLAS uses the total transverse energy, \sumETPb, measured in the negative pseudorapidity sections of the forward calorimeter in the range $-4.9<\eta<-3.1$ (in the direction of the Pb beam)~\cite{Aad:2015zza}. 
The transverse energies in the forward calorimeter are evaluated at an energy scale calibrated for electromagnetic showers and are not corrected for hadronic response~\cite{Aad:2010ad}. 
Centrality intervals are defined in terms of percentiles of the \sumETPb distribution after accounting for an 
estimated inefficiency of approximately $\left(2\pm2\right)$\% for inelastic \pPb events to satisfy the applied event selection criteria. This inefficiency affects mainly the most peripheral events. 
The following centrality intervals are used in this analysis:
0--1\%, 1--5\%, 5--10\%, 10--20\%, 20--30\%, 30--40\%, 40--60\%, 60--90\%  (with the 0--1\% interval defined by the highest \sumETPb\ values). Since the composition of the events and the uncertainty on the inelastic \pPb events selection efficiency in the most peripheral 90--100\% interval is not well constrained, these events are excluded from the analysis, and events from the 60--90\% centrality interval are used as the reference for \rcp.

Following the procedure adopted in Ref.~\cite{Aad:2015zza}, three different estimations of the average number of nucleons participating in the \pPb collisions \avgNpart are carried out in each centrality interval. The first estimation uses the standard Glauber model~\cite{Miller:2007ri}, which is characterized by a fixed total nucleon--nucleon cross section. The other two estimations use the Glauber--Gribov colour-fluctuation (GGCF) model~\cite{Guzey:2005tk,Alvioli:2013vk}, which includes event-by-event fluctuations in the nucleon--nucleon cross section $\sigma_{\mathrm{NN}}~(N+N\rightarrow X)$. In the GGCF model, the magnitude of the fluctuations is characterized by the parameter \ggcf, with \ggcf$=0$ corresponding to the standard Glauber model. Two values, \ggcfA~and~\ggcfB, based on the calculations in Refs. \cite{Guzey:2005tk,Alvioli:2013vk}, are used in this measurement.

In both geometric models the value of \avgTpb is directly related to \avgNpart via the relation 
$\avgNpart - 1 = \avgTpb \sigma_{\mathrm{NN}} $,
with $\sigma_{\mathrm{NN}}$ taken to be $70\pm5$ mb~\cite{Adam:2014qja}.
The obtained \avgTpb values for the Glauber and \GG models 
in different centrality intervals are listed in Table~\ref{tbl:NucTPb}. 
For central collisions, the \avgTpb uncertainties are dominated by the uncertainty in the \Gl/\GG modelling. 
For more peripheral collisions, the uncertainty in the efficiency for selecting inelastic events also makes a significant contribution. 
\begin{table}[htbp]
  \centering
  \renewcommand{\arraystretch}{1.4}
  \sisetup{retain-explicit-plus}
  \sisetup{round-mode = places}
  \sisetup{round-mode = places, round-precision = 1}
  \begin{tabular}{%
      r@{\,/\,}r|
      S[round-mode = places, round-precision = 1,
        table-format = 3.2, table-number-alignment = left]
      @{\,}l@{\,}l |
      S[round-mode = places, round-precision = 1,
        table-format = 3.2, table-number-alignment = left]
      @{\,}l@{\,}l |
      S[round-mode = places, round-precision = 1,
        table-format = 3.2, table-number-alignment = left]
      @{\,}l@{\,}l
    }
    \multicolumn{2}{c|}{Centrality} & \multicolumn{3}{c|}{Glauber} & \multicolumn{6}{c}{\GG} \\
    \multicolumn{2}{c|}{} & \multicolumn{3}{c|}{} & \multicolumn{3}{c|}{\ggcfA} & \multicolumn{3}{c}{\ggcfB} \\ 
    \hline
    \multicolumn{2}{r|}{60--90\%} & 42.3  & $^{+2.8}_{-4.3}$ & $\left( ^{+7\%}_{-10\%}\right)$ & 36.6 & $^{+2.7}_{-2.2}$ & $\left( ^{+7\%}_{-6\%}\right)$ & 34.4 & $^{+3.7}_{-2.1}$ & $\left( ^{+11\%}_{-6\%}\right)$    \\
    \multicolumn{2}{r|}{40--60\%} & 92    & $^{+4}_{-7}$ & $\left( ^{+5\%}_{-7\%}\right)$     & 80.2 & $^{+4.6}_{-3.3}$ & $\left( ^{+6\%}_{-4\%}\right)$ & 75.9 & $^{+6.5}_{-3.3}$ & $\left( ^{+9\%}_{-4\%}\right)$     \\
    \multicolumn{2}{r|}{30--40\%} & 125.6 & $^{+3.3}_{-4.5}$ & $\left( ^{+3\%}_{-4\%}\right)$   & 116.7 & $^{+3.8}_{-3.2}$ & $\left( ^{+3.2\%}_{-2.7\%}\right)$ & 113.1 & $^{+6.6}_{-3.3}$ & $\left( ^{+6\%}_{-3\%}\right)$  \\
    \multicolumn{2}{r|}{20--30\%} & 147.9 & $^{+3.6}_{-2.6}$ & $\left( ^{+2.4\%}_{-1.8\%}\right)$ & 145.5 & $^{+3.6}_{-3.0}$ & $\left( ^{+2.5\%}_{-2.1\%}\right)$& 144.6 & $^{+5.6}_{-3.1}$ & $\left( ^{+4\%}_{-2\%}\right)$   \\
    \multicolumn{2}{r|}{10--20\%} & 172   & $^{+7}_{-3}$ & $\left( ^{+4\%}_{-2\%}\right)$      & 181.9 & $^{+4.4}_{-3.1}$ & $\left( ^{+2.4\%}_{-1.7\%}\right)$ & 186.8 & $^{+5}_{-2.9}$ & $\left( ^{+3\%}_{-2\%}\right)$  \\
    \multicolumn{2}{r|}{5--10\%}  & 194   & $^{+15}_{-4}$ & $\left( ^{+8\%}_{-2\%}\right)$     & 221 & $^{+6}_{-5}$ & $\left( ^{+3\%}_{-2\%}\right)$ & 235 & $^{+7}_{-7}$ & $\left( ^{+3\%}_{-3\%}\right)$   \\
    \multicolumn{2}{r|}{1--5\%}   & 215   & $^{+22}_{-5}$ & $\left( ^{+10\%}_{-2\%}\right)$    & 264 & $^{+9}_{-10}$ & $\left( ^{+3\%}_{-4\%}\right)$ & 292 & $^{+8}_{-23}$ & $\left( ^{+3\%}_{-8\%}\right)$    \\
    \multicolumn{2}{r|}{0--1\%}   & 245   & $^{+40}_{-7}$ & $\left( ^{+15\%}_{-3\%}\right)$   & 330 & $^{+15}_{-23}$ & $\left( ^{+5\%}_{-7\%}\right)$ & 377 & $^{+12}_{-60}$ & $\left( ^{+3\%}_{-16\%}\right)$    \\
    \hline
    \multicolumn{2}{r|}{0--90\%}  & 106.3 & $^{+4.4}_{-2.7}$ & $\left( ^{+4\%}_{-2\%}\right)$ & 107.3 & $^{+3.9}_{-2.6}$ & $\left( ^{+4\%}_{-2\%}\right)$ & 109 & $^{+4}_{-2}$ & $\left( ^{+4\%}_{-2\%}\right)$    \\
  \end{tabular}
  \caption{Mean values of \TPb  in b$^{-1}$ for all centrality intervals, along with asymmetric systematic uncertainties shown as absolute as well as relative uncertainties. The columns correspond to the \Gl\ model (left), \GG\ model with \ggcfA\ (middle), and \GG\ model with \ggcfB\ (right).}
  \label{tbl:NucTPb}
\end{table}

Ratios of the \avgTpb values, which are relevant to \rcp, in a given centrality interval to the respective value in the 60--90\% interval are presented in Table~\ref{tbl:tpbratios}. 
\begin{table}[htbp]
  \centering
  \renewcommand{\arraystretch}{1.4}
  \sisetup{retain-explicit-plus}
  \sisetup{round-mode = places}
  \begin{tabular}{%
      r@{\,/\,}r|
      l@{\,}l@{\,}l|
      l@{\,}l@{\,}l|
      l@{\,}l@{\,}l
    }
    \multicolumn{2}{c|}{Centrality} & \multicolumn{3}{c|}{Glauber} & \multicolumn{6}{c}{\GG} \\
    \multicolumn{2}{c|}{\,} & \multicolumn{3}{c|}{\,} & \multicolumn{3}{c|}{\ggcfA} & \multicolumn{3}{c}{\ggcfB} \\ 
    \hline
    40--60\%&60--90\% & \num[round-precision=2]{2.16} & $^{\numRP{+0.09}{2}}_{\numRP{-0.06}{2}}$ & $\left( ^{+4\%}_{-3\%}\right)$  & \num[round-precision=2]{2.19} & $^{\numRP{+0.04}{2}}_{\numRP{-0.06}{2}}$ & $\left( ^{+2.6\%}_{-2.7\%}\right)$ & \num[round-precision=2]{2.21} & $^{\numRP{+0.05}{2}}_{\numRP{-0.06}{2}}$ & $\left( ^{+2.4\%}_{-2.8\%}\right)$  \\
    30--40\%&60--90\% & \num[round-precision=2]{2.97} & $^{\numRP{+0.22}{2}}_{\numRP{-0.13}{2}}$ & $\left( ^{+7\%}_{-4\%}\right)$  & \num[round-precision=2]{3.19} & $^{\numRP{+0.13}{2}}_{\numRP{-0.13}{2}}$ & $\left( ^{+4\%}_{-4\%}\right)$    & \num[round-precision=2]{3.29} & $^{\numRP{+0.12}{2}}_{\numRP{-0.16}{2}}$ & $\left( ^{+4\%}_{-5\%}\right)$    \\
    20--30\%&60--90\% & \num[round-precision=2]{3.49} & $^{\numRP{+0.34}{2}}_{\numRP{-0.17}{2}}$ & $\left( ^{+10\%}_{-5\%}\right)$ & \num[round-precision=2]{3.98} & $^{\numRP{+0.18}{2}}_{\numRP{-0.21}{2}}$ & $\left( ^{+5\%}_{-5\%}\right)$    & \num[round-precision=2]{4.21} & $^{\numRP{+0.19}{2}}_{\numRP{-0.28}{2}}$ & $\left( ^{+4\%}_{-7\%}\right)$   \\
    10--20\%&60--90\% & \num[round-precision=2]{4.06} & $^{\numRP{+0.5}{2}}_{\numRP{-0.21}{2}}$ & $\left( ^{+13\%}_{-5\%}\right)$ & \num[round-precision=2]{4.98} & $^{\numRP{+0.25}{2}}_{\numRP{-0.31}{2}}$ & $\left( ^{+5\%}_{-6\%}\right)$    & \num[round-precision=2]{5.43} & $^{\numRP{+0.28}{2}}_{\numRP{-0.5}{2}}$ & $\left( ^{+5\%}_{-9\%}\right)$   \\
    5--10\%&60--90\%  & \num[round-precision=2]{4.58} & $^{\numRP{+0.8}{2}}_{\numRP{-0.24}{2}}$ & $\left( ^{+16\%}_{-5\%}\right)$ & \num[round-precision=2]{6.05} & $^{\numRP{+0.33}{2}}_{\numRP{-0.5}{2}}$ & $\left( ^{+5\%}_{-7\%}\right)$     & \num[round-precision=1]{6.8}  &  $^{\numRP{+0.4}{1}}_{\numRP{-0.8}{1}}$  & $\left( ^{+6\%}_{-12\%}\right)$   \\
    1--5\%&60--90\%   & \num[round-precision=2]{5.08} & $^{\numRP{+0.9}{2}}_{\numRP{-0.27}{2}}$ & $\left( ^{+18\%}_{-5\%}\right)$ & \num[round-precision=1]{7.2 } & $^{\numRP{+0.4}{1}}_{\numRP{-0.6}{1}}$   & $\left( ^{+6\%}_{-9\%}\right)$     & \num[round-precision=1]{8.5}  &  $^{\numRP{+0.5}{1}}_{\numRP{-1.4}{1}}$  & $\left( ^{+6\%}_{-16\%}\right)$  \\
    0--1\%&60--90\%   & \num[round-precision=1]{5.8}  & $^{\numRP{+1.3}{1}}_{\numRP{-0.33}{1}}$ & $\left( ^{+23\%}_{-6\%}\right)$  & \num[round-precision=1]{9  }  & $^{\numRP{+0.5}{1}}_{\numRP{-1.1}{1}}$   & $\left( ^{+6\%}_{-12\%}\right)$   & \num[round-precision=1]{11} & $^{\numRP{+0.6}{1}}_{\numRP{-2.6}{1}}$   & $\left( ^{+5\%}_{-23\%}\right)$  \\
  \end{tabular}
  \caption{Ratios of the mean values of \TPb for all centrality bins with respect to the 60--90\% centrality interval, along with the corresponding total systematic uncertainty. The columns correspond to the \Gl model (left), \GG model with \ggcfA (middle), and \GG\ model with \ggcfB (right).
  }
  \label{tbl:tpbratios}
\end{table}

\section{Reconstruction of charged-particle spectra}
\label{sec:recon}
\subsection{Track selection}
\label{sec:trkselect}

The analysis of the charged-particle spectra presented in this paper refers to primary charged particles directly produced in the \pPb or \pp interactions and having a mean lifetime greater than $0.3\times10^{-10}$~s, or long-lived charged particles created by subsequent decays of particles with a shorter lifetime~\cite{Aad:2010ac}. All other particles are considered secondary. Tracks produced by primary and secondary particles are referred to from now on as primary and secondary tracks, respectively.

Tracks are required to be in the kinematic range of transverse momentum $\pT>0.1$~\GeV\ and absolute pseudorapidity $|\eta|<2.3$. Additional requirements on the number of hits in the ID subsystems are imposed in order to reduce the contribution from `fake’ tracks that do not correspond to the passage of charged particles through the detector. All tracks are required to have at least one hit in the pixel detector and a hit in the first pixel layer if one is expected by the track trajectory. Tracks with $\pT<0.2$~\GeV\ are required to have at least two hits in the SCT, tracks with $0.2<\pT<0.3$~\GeV\ are required to have at least four hits in the SCT, and all other tracks are required to have at least six hits in the SCT. To ensure that the tracks originate from the event vertex, the
transverse ($d_0$) and longitudinal ($z_0\sin{\theta}$) impact parameters of the reconstructed track trajectory with respect to the reconstructed primary vertex are required to be less than 1.5~mm. Finally, tracks are required to satisfy the significance conditions $|d_0/\sigma_{d_0}|<3.0$ and $|z_0\sin{\theta}/\sigma_{z_{0}\sin{\theta}}|<3.0$, where the quantities $\sigma_{d_0}$ and $\sigma_{z_{0}\sin{\theta}}$ are the uncertainties in the determination of $d_0$ and $z_0\sin{\theta}$ obtained from the covariance matrix provided by the ATLAS track model~\cite{Aad:2010ac}.

In \pp collisions, tracks  originating from all reconstructed vertices are used in the analysis. 
The track-to-vertex matching uses the track $z_{0}$ parameter and the $z$ coordinate of the vertex. 
These parameters of the tracks in \pp collisions are often less precisely defined than in \pPb due to the fact that the vertices are typically reconstructed with fewer tracks. Thus in the \pp data analysis the track selection cuts related to the vertex are relaxed such that the $z_0\sin{\theta}$ impact parameter condition is required to be less than 2.5~mm and the transverse and longitudinal impact parameter significances are required to be less than 4.0.

For the calculation of \rpPb, the momentum three-vector is used to calculate the rapidity of the particle, assuming it has the mass of the pion (\mpi). A correction for this assumption is discussed in Section~\ref{sec:trkcorr}. 

\subsection{Reconstruction of the invariant particle distributions}
\label{sec:trkcorr}
The per-event \pPb charged-particle multiplicity distributions are measured differentially as a function of \pT and either $\eta$ or \ystar, and are referred to as the differential invariant yields. They are defined as:
\begin{eqnarray}
\frac{1}{N_{\mathrm{evt}}}\frac{1}{2\pi\pT}\frac{\mathrm{d}^{2}N_{\mathrm{ch}}}{\mathrm{d}\pT d\eta} = 
\frac{1}{N_{\mathrm{evt}}}\frac{1}{2\pi\pT} 
\frac{N_{\mathrm{ch}}(\pt,\eta)}{\Delta\pT \Delta\eta} \frac{{\cal P}(\pT,\eta)}{{\cal C}_{\mathrm{trk}}(\pT,\eta)} 
\quad {\textrm and} \quad\\
\frac{1}{N_{\mathrm{evt}}}\frac{1}{2\pi\pT}\frac{\mathrm{d}^{2}N_{\mathrm{ch}}}{\mathrm{d}\pT d\ystar} = 
\frac{1}{N_{\mathrm{evt}}}\frac{1}{2\pi\pT} 
\frac{N_{\mathrm{ch}}(\pt,\ystar\,)}{\Delta\pT \Delta\ystar} \frac{{\cal P}(\pT,\eta){\cal A}(\pt,\ystar\,)}{{\cal C}_{\mathrm{trk}}(\pT,\eta)}, 
\label{eq:Eff_ycms}
\end{eqnarray}
where $\Delta\pT$, $\Delta\eta$ and $\Delta\ystar$ are the widths of the transverse momentum, pseudorapidity
and rapidity intervals being studied, and $N_{\mathrm{evt}}$ is the number of events in the analysed centrality interval. 
The correction factors ${\cal C}_{\text{trk}}$, ${\cal P}$, and ${\cal A}$ are used to correct for track efficiency and transverse momentum resolution, contributions from fake tracks and secondaries, and to transform the distributions from \ypi to hadron rapidity, respectively.

The correction factor used to correct for the track reconstruction inefficiency is estimated from simulation and is defined as: 
\begin{eqnarray}
{\cal C}_{\mathrm{trk}}(\pT,\eta) = \frac{N^{\mathrm{Rec}}_{\mathrm{Primary}}(\pT,\eta)}{N^{\mathrm{Gen}}_{\mathrm{Primary}}(\pT^{\mathrm{Gen}},\eta^{\mathrm{Gen}})},
\label{eq:Eff_eta}
\end{eqnarray}
where $N^{\mathrm{Gen}}_{\mathrm{Primary}}$ is the number of primary charged particles and $N^{\mathrm{Rec}}_{\mathrm{Primary}}$ is the number of reconstructed tracks that are matched to those charged particles. A track is matched to a generated particle if that particle contributes more than $50$\% to the weighted number of hits on the track. The hits are weighted such that all subdetectors have the same weight in the sum. 
The algorithm to match reconstructed tracks to generated particles is discussed in Ref.~\cite{Cornelissen:1092934}. 
These correction factors are calculated using Monte Carlo events generated with the $\HIJING$ event generator. 
The correction factors are calculated after reweighting the particle-level spectra to achieve better agreement 
in the transverse momentum distribution between data and simulation. 
The track reconstruction correction factor values are smaller at low \pT,
starting at around 20\% in the lowest measured interval of $0.1<\pT<0.2$~\GeV, and then increase rapidly to reach a plateau value
at approximately 1~\GeV. The plateau of the correction factor values is generally higher in the centre of the detector,
reaching 80\% for highest \pT and $\eta=0$, but only 60\% at $|\eta|=2.3$. 
This correction has a very weak centrality dependence; the maximum variation from peripheral to central
collisions does not exceed 2\% over the range of measured
centralities at any \pT or $\eta$ value.

The correction factors to remove the contributions from fake and residual secondary tracks are estimated from simulation and are given by:
\begin{equation}
  {\cal P}(\pt,\eta) = \frac{N_{\mathrm{Primary}}^{\mathrm{Rec}}(\pT,\eta)}{N^{\mathrm{Rec}}(\pT,\eta)}, 
  \label{eq:frac}
\end{equation}
where $N^{\mathrm{Rec}}$ is the total number of reconstructed particles. 
This correction has a strong dependence on both \eta\ and \pT\ at the lowest transverse momentum. 
The value of ${\cal P}$ is 0.98 for tracks with $\pT>1$~\GeV\ in all \eta\ and centrality intervals, dropping to 0.8 for tracks at $|\eta|\sim2.3$ in the 0--1\% centrality interval. 

\begin{figure}[!htb]
\begin{center}
\includegraphics[width=0.59\textwidth]{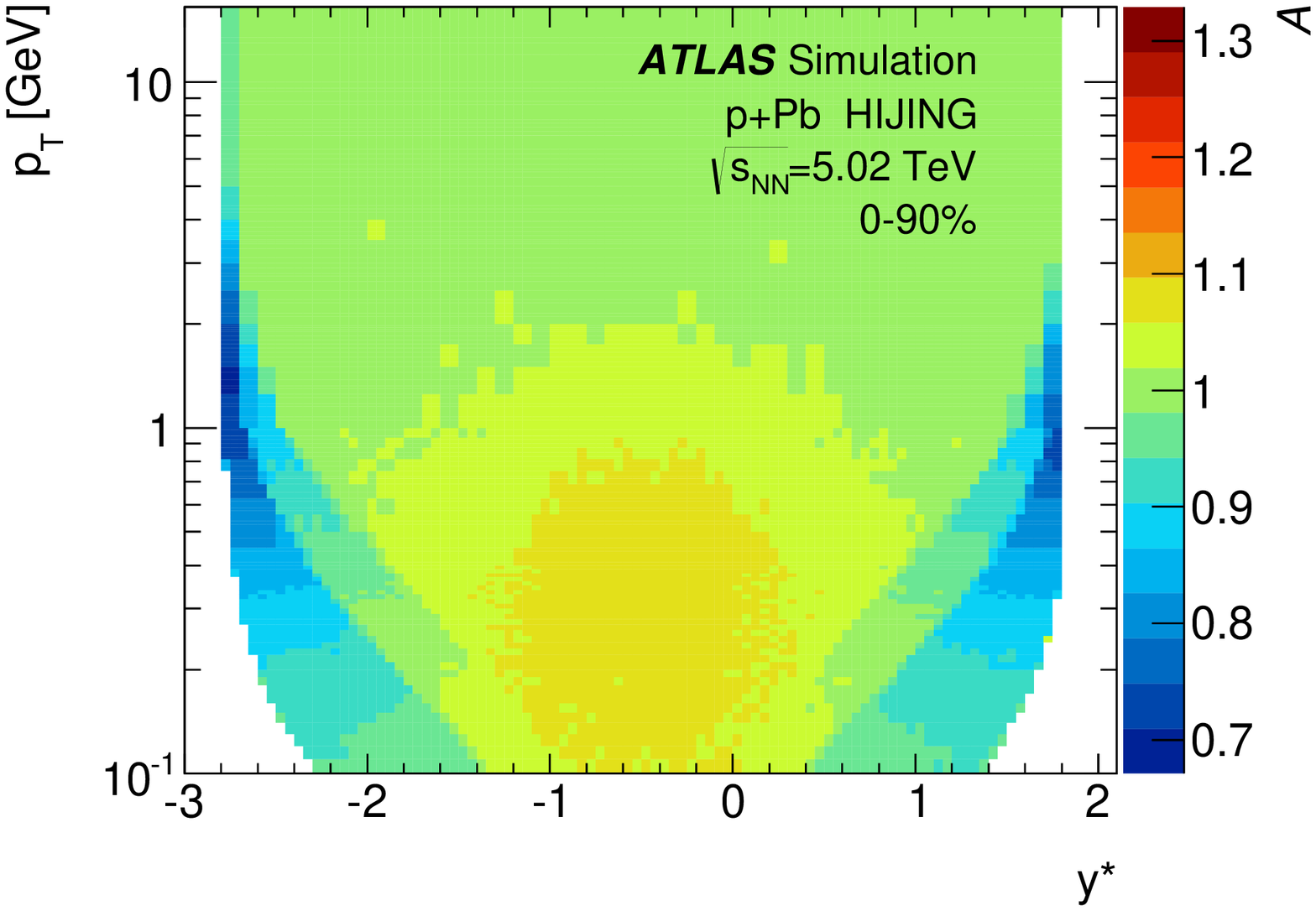}
\caption[Correction.]{${\cal A}$(\pT,\ystar) as a function of \pT and \ystar for the \pPb MC sample.} 
\label{fig:corr_pKpi_pPb}
\end{center}
\end{figure}

The assumption that the particle mass is equal to the pion mass is used to calculate \ystar from the track's momentum three-vector. For tracks that are not pions, the \ystar is computed incorrectly and the particle contributes to the yield in the wrong \ystar bin. A correction for this effect is derived from the simulation as the ratio in \pT and \ystar space of the generated charged particles with their correct mass to the corresponding distribution of generated charged particles assumed to be pions:
\begin{equation}
  \left. {\cal A}(\pt,\ystar) = \frac{N^{\mathrm{Gen}}_{\mathrm{Primary}}(m,\pT,\ystar)}{N^{\mathrm{Gen}}_{\mathrm{Primary}}(m_{\pi},\pT,\ystar)} \right. .
  \label{eq:corr}
\end{equation}
The correction function is shown in Figure~\ref{fig:corr_pKpi_pPb} as a two-dimensional distribution for \pt\ and \ystar\ in the \pPb\ system. 
The correction is approximately 1.1 at $\ystar=0$ and decreases to unity with increasing \pT, as the influence of the mass of the particle on the rapidity becomes negligible. At the edges of acceptance ($\ystar\approx-2.3$), the value of ${\cal A}$ is approximately 0.8 for particles with $\pT\approx 0.7$~\GeV. Fiducial regions with ${\cal A}\leq 0.9$ are removed from the analysis of \rpPb, using the selection criteria documented in Table~\ref{tab:fidutial_cut}. 
This ensures minimal model dependence in the correction factor.

\begin{table}[ht]
\centering      
\begin{tabular}{c|ccc}
\pt\ range [\gev]    & \pPb\ \ystar\ range  & \pp\ \ystar\ range  & Combined \ystar\ range\\
\hline
{$\!\begin{aligned}
0.1<&\pt<0.4 \\      
0.4<&\pt<1 \\         
 1<&\pt<2 \\           
 2<&\pt<3 \\           
 &\pt>3
\end{aligned}$}
& 
{$\!\begin{aligned}            
-2.3<&\ystar<1.3 \\   
-2.5<&\ystar<1.5 \\  
-2.7<&\ystar<1.7 \\   
-2.75<&\ystar<1.75 \\ 
-2.8<&\ystar<1.8 \\  
\end{aligned}$}
&
{$\!\begin{aligned}            
-1.8<&\ystar<1.8    \\   
-2.0<&\ystar<2.0      \\ 
-2.2<&\ystar<2.2    \\   
-2.25<&\ystar<2.25  \\    
-2.3<&\ystar<2.3    \\ 
\end{aligned}$}
&
{$\!\begin{aligned}            
-1.8<&\ystar<1.3  \\
-2.0<&\ystar<1.5  \\
-2.2<&\ystar<1.7  \\
-2.25<&\ystar<1.75  \\
-2.3<&\ystar<1.8  \\
\end{aligned}$}

\end{tabular}
\caption{Fiducial cuts for the combination of \pPb\ and \pp\ acceptance effects.}
\label{tab:fidutial_cut}
\end{table}

\section{Reference spectra from \pp collisions}
\label{sec:ppref}

The differential charged-particle cross sections for \pp\ collisions are defined in an analogous way to those used for \pPb\ differential invariant yield by:
\begin{eqnarray}
\frac{1}{2\pi\pT}\frac{\mathrm{d^{2}}\mathrm{\sigma}_{\footnotesize\pp}}{\mathrm{d}\pT d\ystar} = 
\frac{1}{2\pi\pT}\frac{1}{L} 
\frac{N_{\mathrm{ch}}(\pt,\ystar\,)}{\Delta\pT \Delta\ystar} \frac{{\cal P}(\pT,\eta){\cal A}(\pt,\ystar\,)}{{\cal C}_{\mathrm{trk}}(\pT,\eta)}, 
\label{eq:Eff_ycms}
\end{eqnarray}
where $L$ is the integrated luminosity of the dataset under consideration. 
The values of ${\cal C}_{\text{trk}}$, ${\cal P}$, and ${\cal A}$ are calculated using MC events produced by the  $\PYTHIA6$ event generator. 
The trigger and vertex reconstruction efficiency in \pp\ data analysis is estimated in Ref.~\cite{Aad:2010ac} to be close to unity and is therefore not corrected for in the analysis (the systematic uncertainty due to this choice is discussed in Section~\ref{sec:syst}). 

\begin{figure}[!ht]
\begin{center}
  \subfloat[\label{subfig-1:interpol}]{%
    \includegraphics[width=0.49\textwidth]{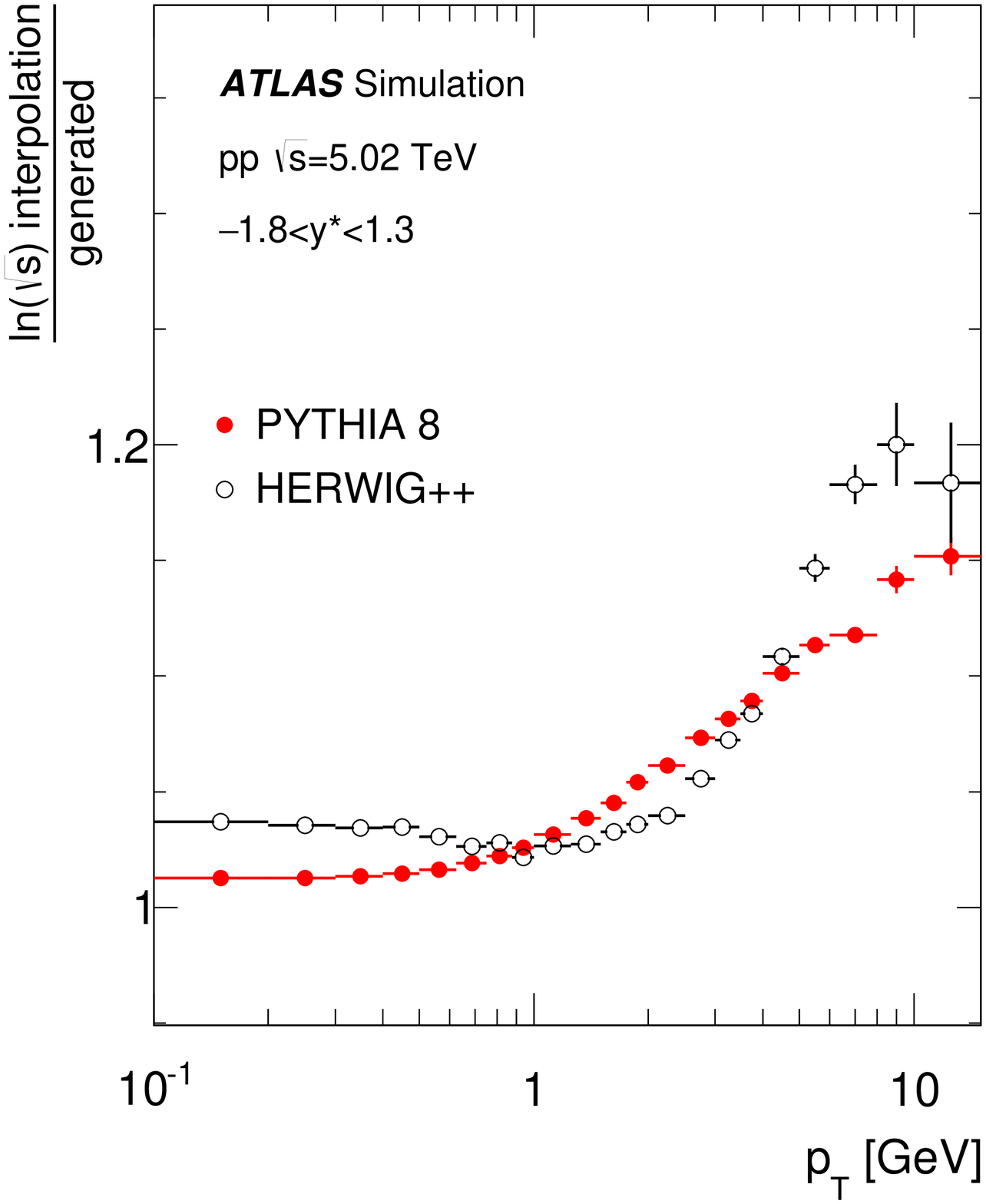}
  }
  ~
  \subfloat[\label{subfig-2:interpol}]{%
    \includegraphics[width=0.49\textwidth]{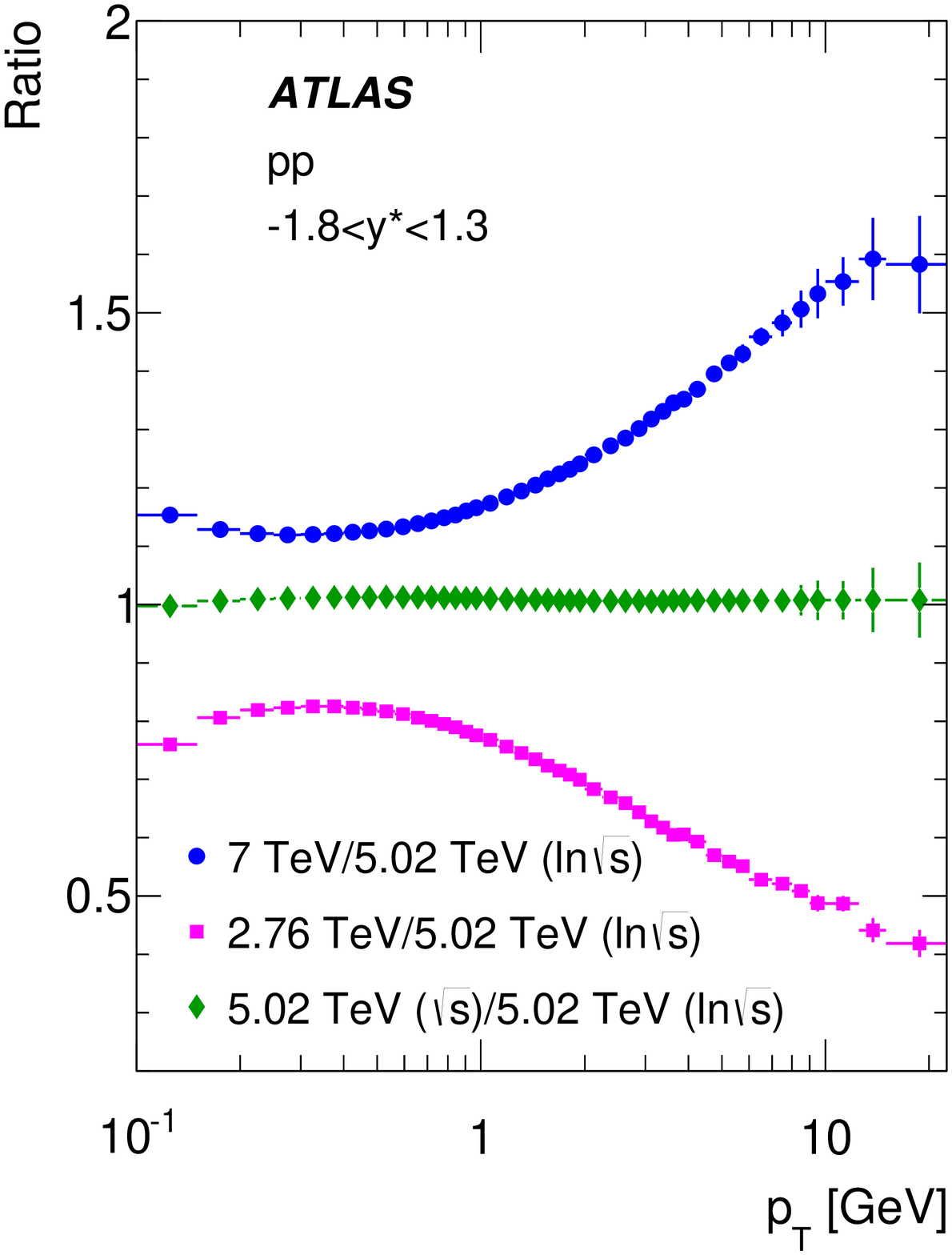}
  }
\caption{(a) The correction factors that are applied to the data. They are obtained as a ratio of the simulated differential cross section at $\sqs=5.02$~\TeV\ to the interpolated cross section, obtained from simulated samples at $\sqs=7$~\TeV\ and $\sqs=2.76$~\TeV\ with $\PYTHIA8$ and \HERWIGpp. (b) The ratios of the input invariant cross sections at $\sqs=7$~\TeV\ (blue circles) and at $\sqs=2.76$~\TeV\ (magenta squares) to the interpolated cross section at $\sqn=5.02$~\TeV. The error bars represent the statistical uncertainties of the input spectra. The comparison between interpolation using \sqs\ and \lns\ is shown with green diamond markers. All the ratios are extracted within the maximal acceptance of the ID detector ($-1.8<\ystar<1.3$).}
\label{fig:interpol}
\end{center}
\end{figure}

Once the differential cross sections at 2.76 and 7~\TeV\ are measured, the charged-particle cross section at $\sqs=5.02$~\TeV\ is estimated by interpolation. Two interpolation functions are investigated for every \pT\ bin in each rapidity interval. The first function is proportional to \sqs, and the second is proportional to \lns. 
The \lns-based interpolation is taken as the default in the analysis and the \sqs-based interpolation is used to assess the systematic uncertainty due to the choice of interpolation function. 
Possible distortions introduced by the interpolation algorithm are evaluated using MC simulations based on $\PYTHIA8$. 
The ratio of the simulated differential cross section at $\sqs=5.02$~\TeV\ to the cross section interpolated with \lns-based or \sqs-based function, obtained from simulated samples at $\sqs=7$~\TeV\ and $\sqs=2.76$~\TeV, is taken as a multiplicative correction factor to be applied to the data. The correction factors obtained using $\PYTHIA8$ and \HERWIGpp are presented in Figure~\ref{subfig-1:interpol} for the region $-1.8<\ystar<1.3$. The correction obtained from $\PYTHIA8$ is the default applied to the data and the correction obtained using \HERWIGpp is used to assess the systematic uncertainty as discussed in Section~\ref{sec:syst}, and calculated separately for either the\lns-based or \sqs-based interpolation functions.

Figure~\ref{subfig-2:interpol} summarizes the relative shapes of the differential cross sections measured at $\sqs=2.76$, $7$ and $5.02$~\TeV, with the last obtained by interpolation. It shows that the effect of the interpolation on the input cross section at $\sqs=2.76$~\TeV\ ($\sqs=7$~\TeV) compared to the interpolated cross section at $\sqs=5.02$~\TeV, using \lns, is 0.8 (1.1) at low \pT values and is 0.4 (1.6) at the highest transverse momentum. The ratio of \sqs-based interpolation to the default \lns-based interpolation shown in the Figure~\ref{subfig-2:interpol} is one of the systematic uncertainties in the cross section interpolation, which are discussed in Section~\ref{sec:syst}.

\section{Systematic uncertainties}
\label{sec:syst}

The systematic uncertainties in the measurement of invariant charged-particle yields arise from inaccuracies of 
the detector description in the simulation, sensitivity to selection criteria used in the analysis and differences between 
the composition of particle species in the simulation and in the data samples. 
To evaluate each source of uncertainty, 
each parameter used in the analysis, such as the values of the quantities used in the track selection criteria or simulated particle composition, is altered within appropriate limits, as described below. 
All sources of systematic uncertainty are evaluated independently in terms of 
$\eta$ and \ystar.

The uncertainty due to the track selection is sensitive to possible differences in performance of the track reconstruction algorithms in data and in MC simulation. To estimate this uncertainty, the basic requirements on the number of detector hits and the track impact parameters were relaxed and tightened in both data and MC simulation. For the relaxed criteria the $d_0$ and $z_0\sin{\theta}$ impact parameters for \pPb~(\pp) samples are required to be less than 2~mm~(3~mm) and significance conditions are not required. To tighten the selection, tracks are required to have at least seven SCT hits, traverse an active module in each layer of the pixel detector, and the impact parameter requirement is changed to be less than 1~mm and 2~mm for \pPb and \pp samples respectively. These variations produce up to a 2\% shift in the fully corrected charged-particle yield. 
The uncertainty in the charged-particle yield due to simulation of inactive material is estimated using dedicated \pPb\ simulated samples in which the inactive material is increased 
in the central and forward regions of the inner detector~\cite{Aad:2011cxa}. 
The net effect on the per-event charged-particle yields is found to vary from 0.5\% at low pseudorapidity to 4\% at high pseudorapidity, but is independent of centrality. 
The systematic uncertainty estimated in this way from \pPb\ simulated samples is applied to both \pPb and \pp data, taking into account the rapidity boost.

The correction for track reconstruction inefficiency, secondaries and fake tracks is calculated from simulated samples after reweighting the track \pT and $\eta$ distributions to match those observed in data. 
The systematic uncertainty in this procedure is derived by taking the difference 
between the results obtained with reweighting and without reweighting of the simulation.

Our imperfect knowledge of the particle composition in \pPb collisions is a source of systematic uncertainty, which influences ${\cal A}$(\pT,\ystar) for $\eta\rightarrow\ystar$ transformation. To assess the sensitivity of the analysis to the particle composition in the $\HIJING$ samples used to correct the data, the relative contributions of the pions, kaons and protons in $\HIJING$ were reweighted to match the fractions obtained from the identified particle multiplicity measured by the ALICE experiment~\cite{Abelev:2013haa}. 
The weights of the charged-particle yields vary from $0.5$ to $1.5$ at low \pT\ and high \pT\ respectively, increase with centrality, and do not depend on $\eta$. 
The change in the charged-particle yields is found to be between $4$\% and $0.1$\% at low \pT\ and high \pT\ respectively, but the variation does not depend on $\eta$.
Variation of the particle composition results in a maximum $5$\% difference in the fully corrected charged-particle yields at moderate and high \ystar\ and low \pT. 
The difference decreases with \pT\ and depends on \ystar, reaching minimum values close to $\ystar=-2$ and $1$. For the \pp analysis, the \pPb multiplicity measurement by the ALICE experiment for the peripheral centrality interval was adopted to estimate the weights. The change in the charged-particle yields is found to be between $2$\% and $0.1$\% at low \pT\ and high \pT\ respectively, and the variation does not depend on $\eta$ and \ystar.

The uncertainties associated with the centrality selection contain the effects of the trigger and event selection criteria. Using the procedure outlined in Ref.~\cite{Aad:2015zza}, the centrality intervals are redefined after assuming a total event selection efficiency, differing by $\pm2$\% from the nominal one, and the change in the multiplicity spectrum reconstructed in each centrality interval is taken as a systematic uncertainty associated with the centrality determination.

In the \pp\ data analysis, the systematic uncertainty assigned to the trigger efficiency is 1\% for events containing two tracks and decreases rapidly with higher
track multiplicities. A transverse momentum and rapidity independent uncertainty of 0.5\% is assigned to the differential cross sections. In the same way as for the trigger efficiency, 
the uncertainty in the vertex reconstruction efficiency in the \pp\ data analysis is taken to be 1\%~\cite{Aad:2010ac}.
   
The systematic uncertainty in the interpolated \pp\ cross section is needed for the correction applied to the interpolated data derived from simulated samples. 
The systematic uncertainty is taken to be the difference between the corrections obtained from $\PYTHIA8$ and \HERWIGpp, which are shown in Figure~\ref{subfig-1:interpol}. An additional systematic uncertainty is estimated by considering the relative difference between spectra obtained using the two different interpolation functions (\sqs or \lns) as shown in Figure~\ref{subfig-2:interpol}.
 
\begin{table}[h]
\begin{center}
\begin{tabular}{l|cc|l}
Uncertainty                    & \pPb       & \pp               & Variation \\
\hline			      				            	 
Track selection                & 2\%        & 1\%               & decreases with \pT, increases with $|\eta|$ \\
Particle composition           & 1--5\%     & 1--2\%            & changes with \pT and \ystar \\
Material budget                & \multicolumn{2}{c|}{0.5--4\%}  & decreases with \pT, increases with $|\eta|$      \\
\pT\ reweighting               & 0.1--0.5\% & 0.1--2.5\%        & decreases with \pT, increases with $\eta$                             \\
Centrality selection           & 0.1--8\%   &    --             & increases with \pT and asymmetric in $\eta$, \\
                               &            &                   & increases with centrality interval width \\ 
Trigger Efficiency             & 0.01\%     & 0.5\%             &                   \\
Luminosity                     & --         & 2.7\% (1.8\%)     & $\sqs=2.76$~\TeV\ (7~\TeV) \\
\pp\ reference interpolation   &  --        & 0.1--5\%          & increases with \pT and constant in $\eta$      \\
Vertex reconstruction          &  0.1\%     &     1\%           &                   	     \\
\end{tabular}
\end{center}
\caption{Systematic uncertainties on charged-particle yields for \pPb and \pp at 2.76~\TeV.
The uncertainty in the luminosity does not contribute to the \pPb results, since they are
expressed as per-event invariant yields.  The uncertainty in the trigger and event selection is 
included in the uncertainty in the efficiency for selecting inelastic events, and thus 
is already contained in the centrality selection's uncertainties.}
\label{tab:pPb_pp_syst}
\end{table}
The uncertainties in the calculated luminosity values for the corresponding \pp\ data samples at $\sqs=7$~\TeV\ and $\sqs=2.76$~\TeV\ are 1.8\%~\cite{Aad:2013ucp} and 2.7\%~\cite{Aad:2013lpa}, respectively. 
They are taken to be fully uncorrelated, thus the total uncertainty in the interpolated spectra at $\sqs=5.02$~\TeV\ is obtained by adding in quadrature the luminosity uncertainties of the inputs. 

A summary of the systematic uncertainties in the charged-particle invariant yields in \pPb and \pp data analysis is shown in Table~\ref{tab:pPb_pp_syst}. For \rpPb and \rcp, some of the errors are correlated between numerator and denominator. Track selection, particle composition, reweighting, trigger efficiency and vertex reconstruction uncertainties largely cancel for \rcp, since the corrections do not vary with centrality interval and the yields are compared in the same \pT and $\eta$ bins. However, for \rpPb, there is little cancellation between \pPb and \pp, since the results are presented as a function of \ystar\ and the two systems are in two different centre-of-mass frames. The systematic uncertainties in \avgTpb and their ratios which are presented in Tables~\ref{tbl:NucTPb}~and~\ref{tbl:tpbratios} are added in quadrature to the experimental uncertainties of \rpPb and \rcp respectively.

\section{Results} 
\label{sec:results}
\begin{figure}[h]
\begin{center}
  \subfloat[\label{subfig-1:pt_1pad_6EtaBins_G}]{%
  \includegraphics[width=0.49\textwidth]{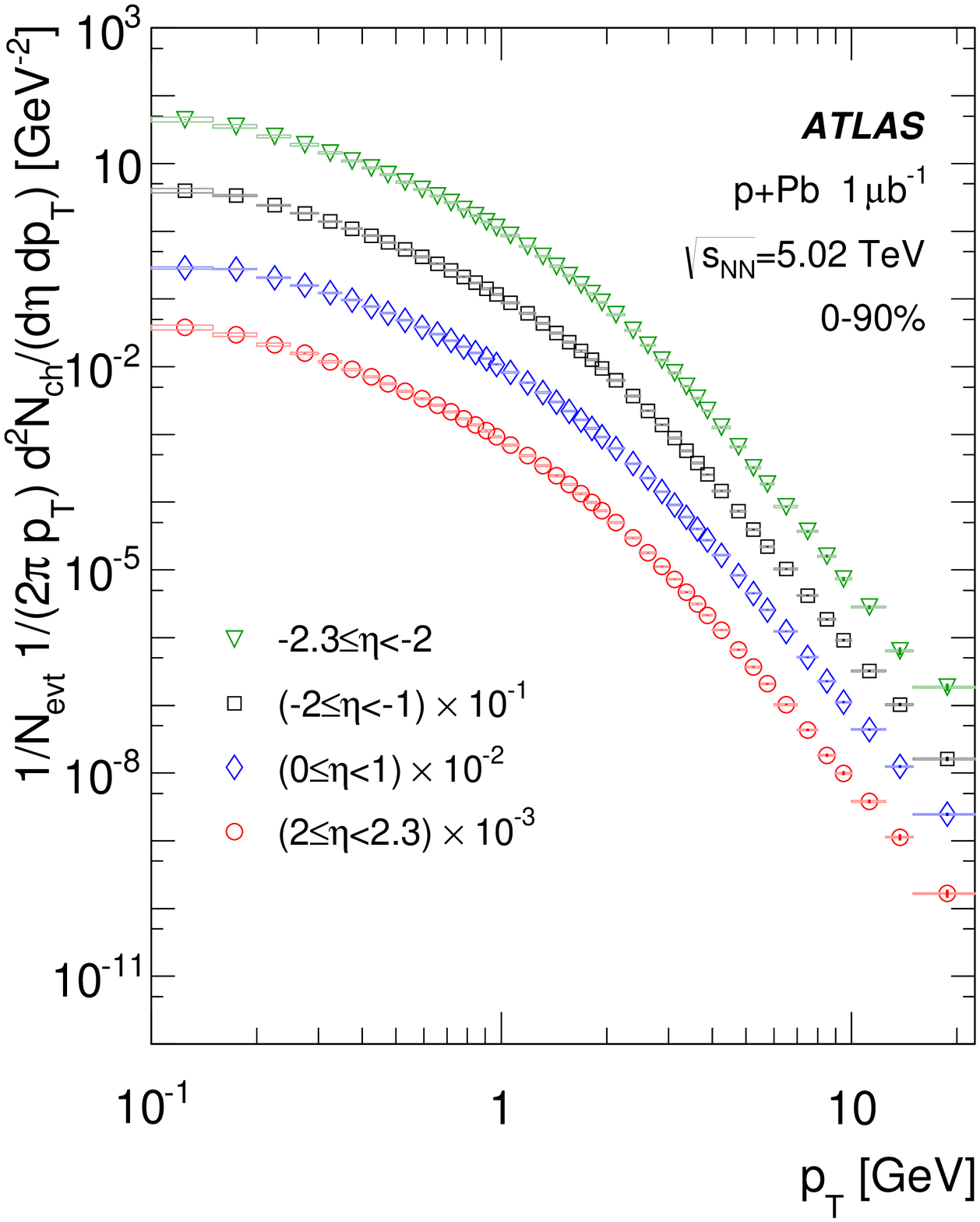}
  }    
  ~
  \subfloat[\label{subfig-2:pt_1pad_6EtaBins_G}]{%
  \includegraphics[width=0.49\textwidth]{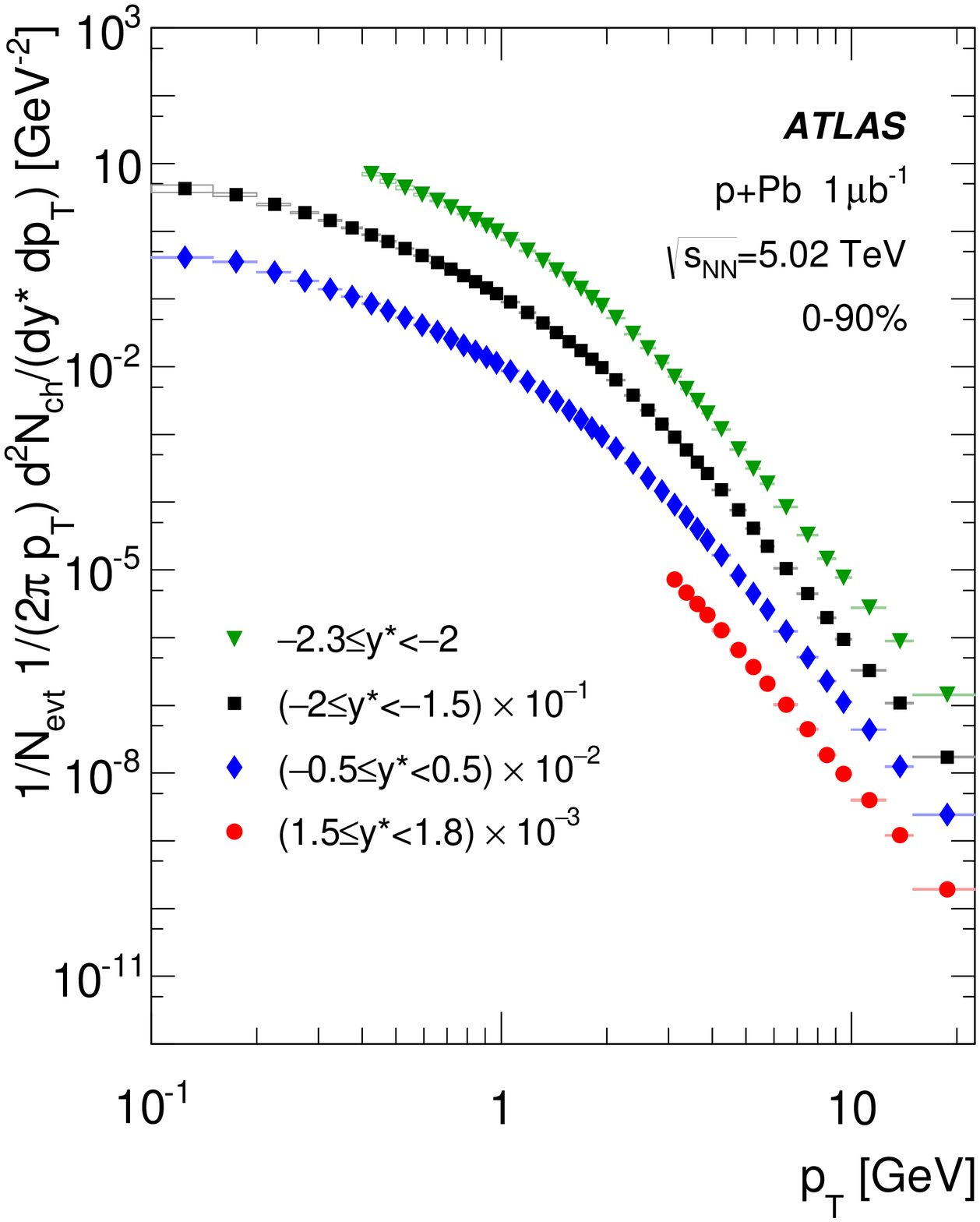}
}
\caption{Invariant differential \pT spectra of charged particles which are produced in \pPb collisions at $\sqs=5.02$~\TeV\ shown in (a) four $\eta$ intervals and (b) four \ystar intervals, for the 0--90\% centrality interval. The individual spectra are scaled by constant factors (indicated in the legend) for visibility. The statistical uncertainties are indicated with vertical lines and the systematic uncertainties are indicated with boxes, but are generally much smaller than the size of the symbols.
} 
\label{fig:pt_1pad_6EtaBins_G}
\end{center}
\end{figure}

The differential invariant yields of charged particles produced in \pPb\ collisions at $\sqs=5.02$~\TeV\ are presented as a function of charged-particle transverse momentum in Fig.~\ref{fig:pt_1pad_6EtaBins_G} for several intervals of $\eta$ and \ystar. 

\begin{figure}[h]
\begin{center}
\includegraphics[width=0.43\textwidth]{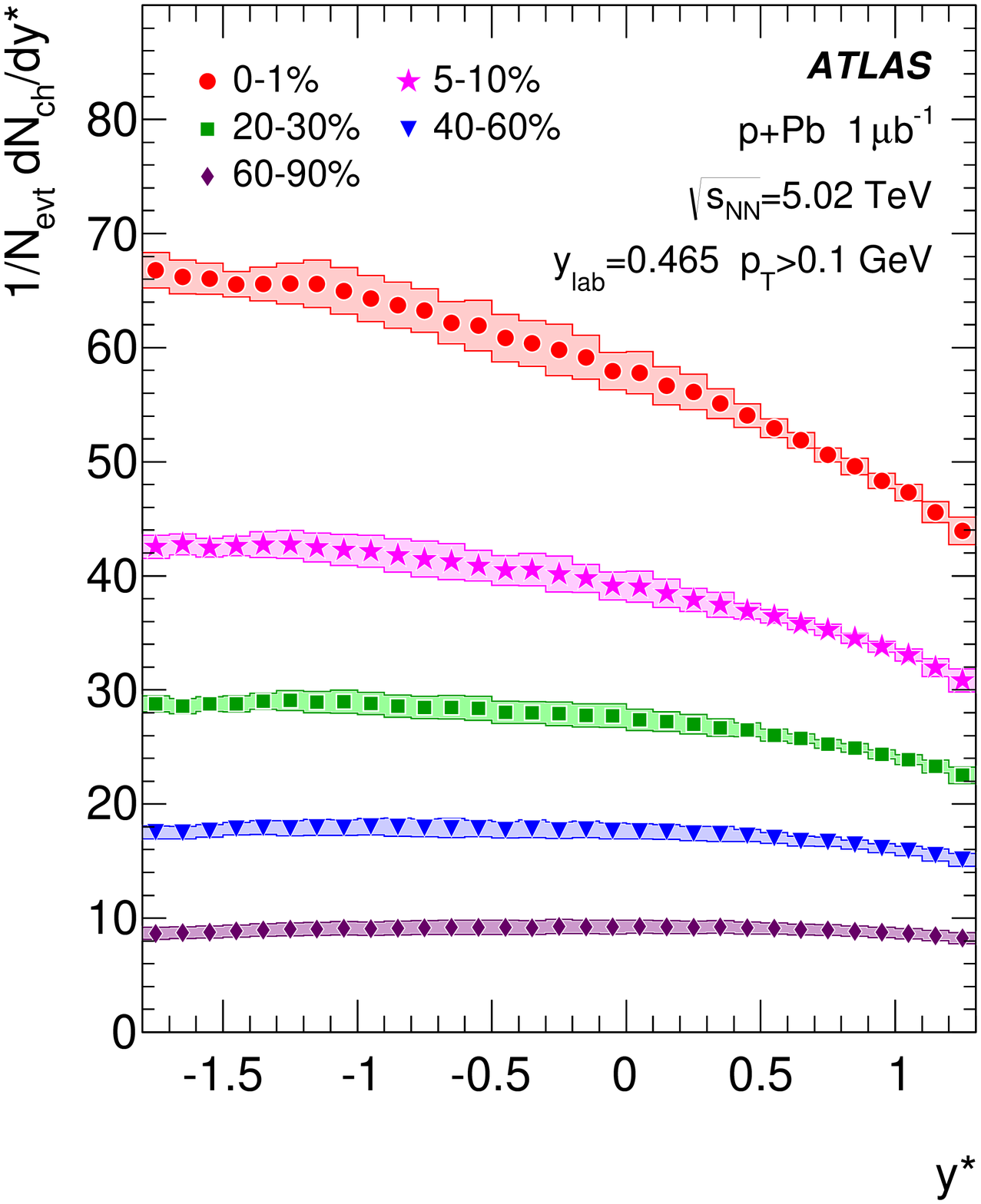}
\caption{The invariant differential \ystar\ spectra of charged particles produced in \pPb collisions at $\sqs=5.02$~\TeV\ are shown in five centrality intervals for $\pT>0.1$~\GeV. The statistical uncertainties are indicated with vertical lines and the systematic uncertainties are indicated with boxes.} 
\label{fig:y_spectra}
\end{center}
\end{figure}

Figure~\ref{fig:y_spectra} shows the invariant charged-particle yield as a function
of \ystar for $\pT>0.1$~\GeV\ in several centrality intervals. 
In collisions that are more central, the charged-particle yields become progressively more asymmetric, as shown in the ATLAS multiplicity measurement~\cite{Aad:2015zza}, with more particles produced in the Pb-going direction than in the proton-going direction.

\begin{figure}[b!]
\begin{center}
  \subfloat[\label{subfig-1:RpPb_0_90_G}]{%
        \includegraphics[width=0.33\textwidth]{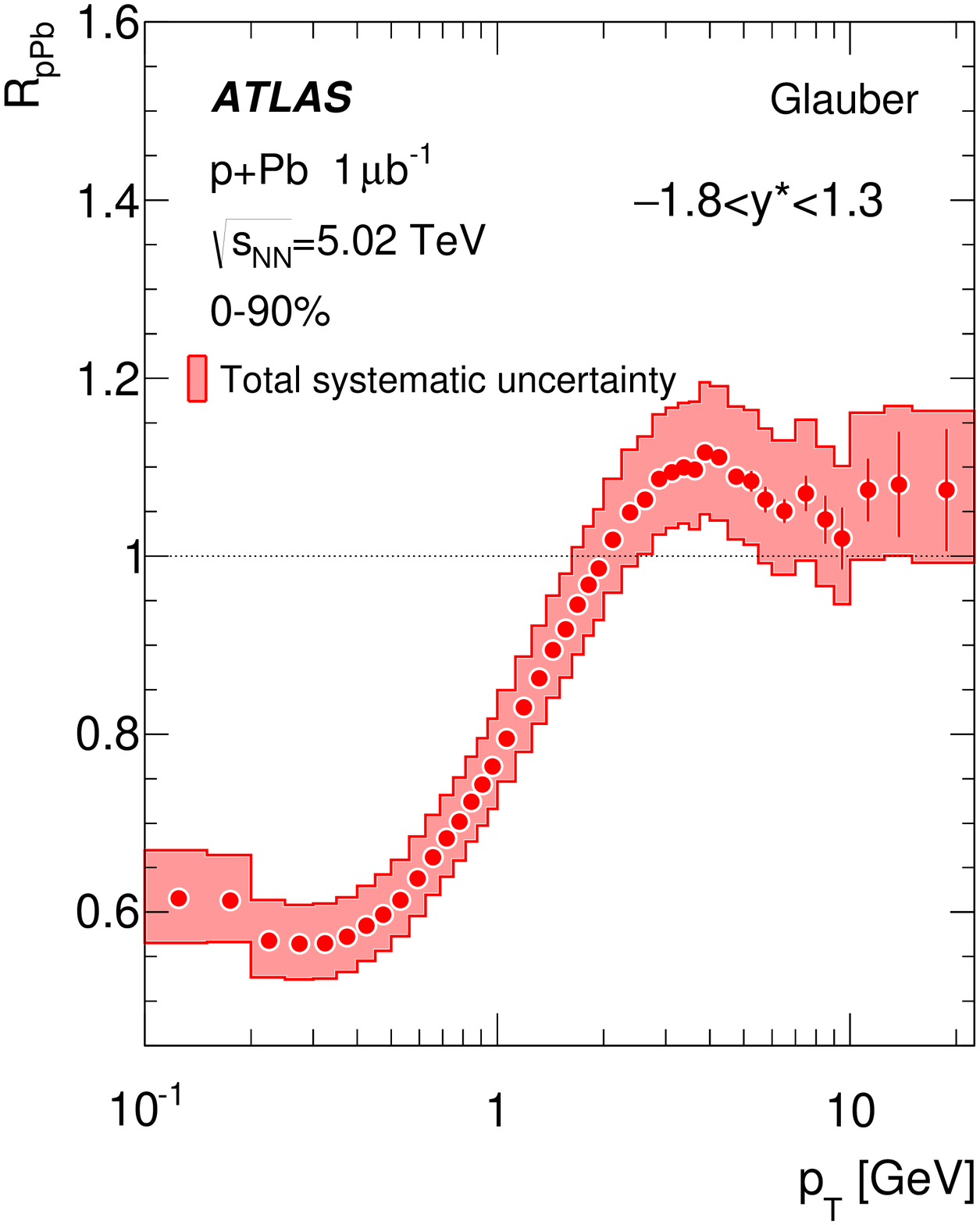}
        }
        ~
  \subfloat[\label{subfig-2:RpPb_0_90_G}]{%
        \includegraphics[width=0.33\textwidth]{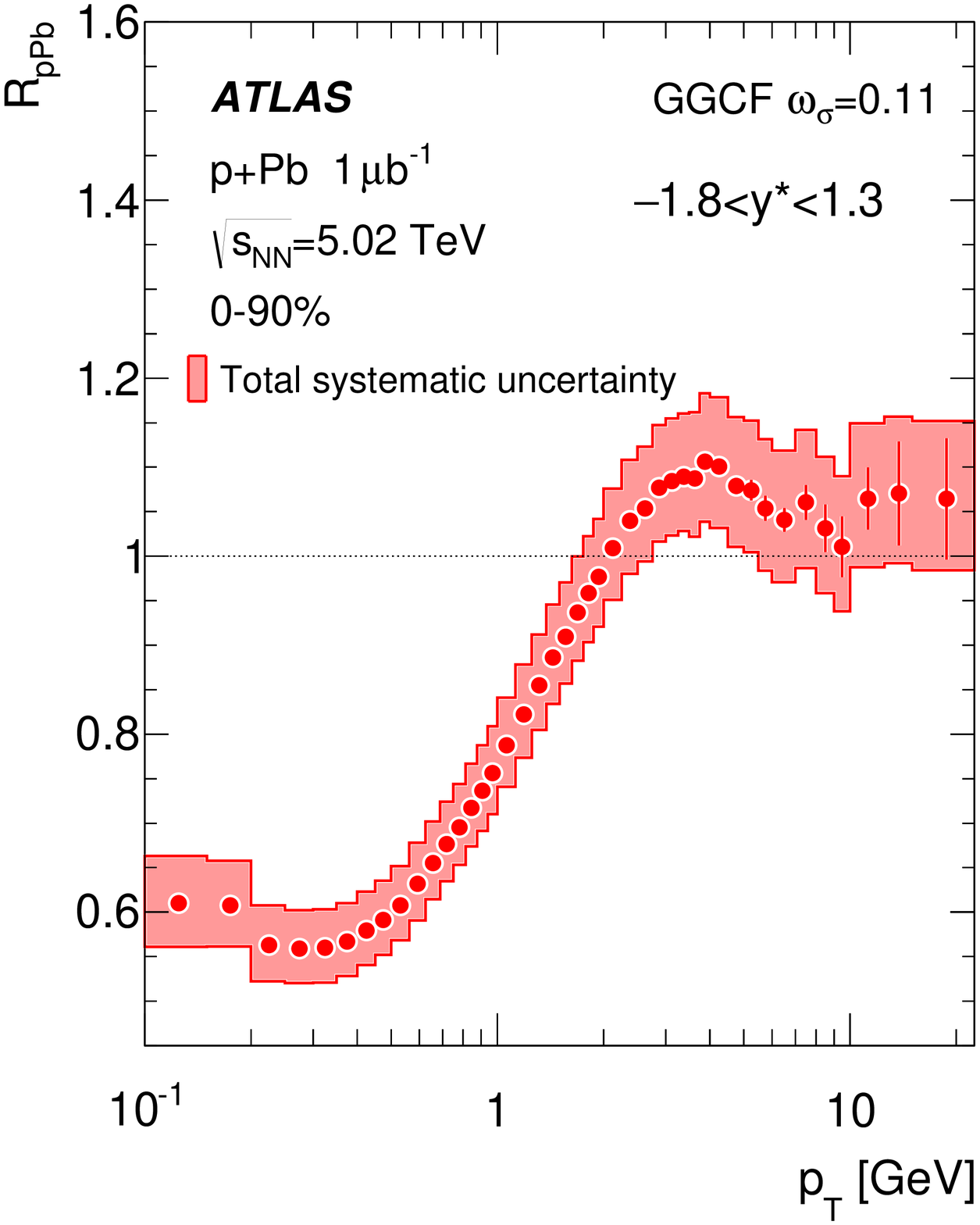}
        }
        ~
  \subfloat[\label{subfig-3:RpPb_0_90_G}]{%
        \includegraphics[width=0.33\textwidth]{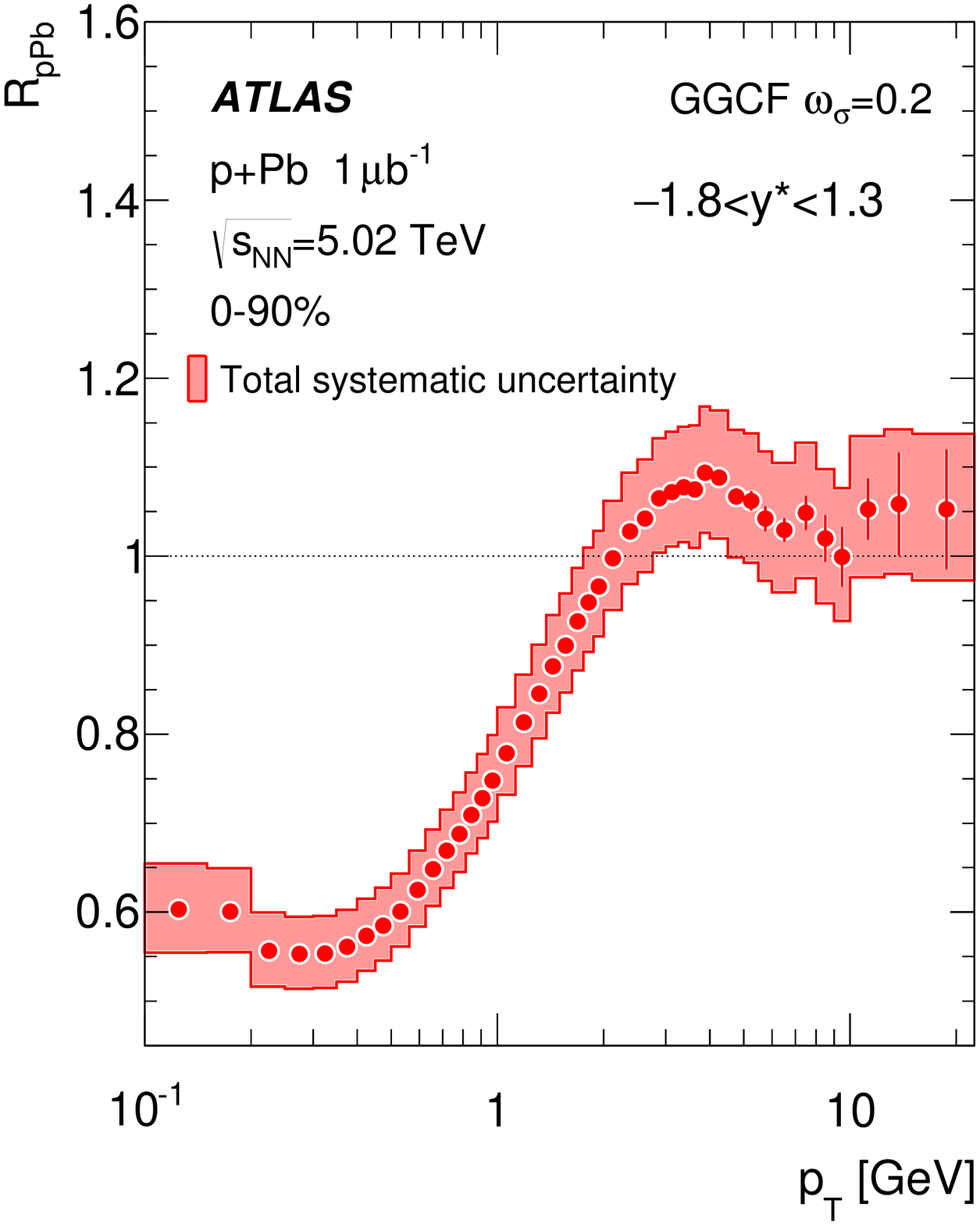}
        }
\caption{\rpPb as a function of \pT\ integrated over rapidity range $-1.8<\ystar<1.3$ for the 0--90\% centrality interval for the three geometric models: (a) \Gl, (b) \GG\ with \ggcfA\ and (c) \GG with \ggcfB.} 
\label{fig:RpPb_0_90_G}
\end{center}
\end{figure}

\begin{figure}[b!]
\begin{center}
\includegraphics[width=0.47\textwidth]{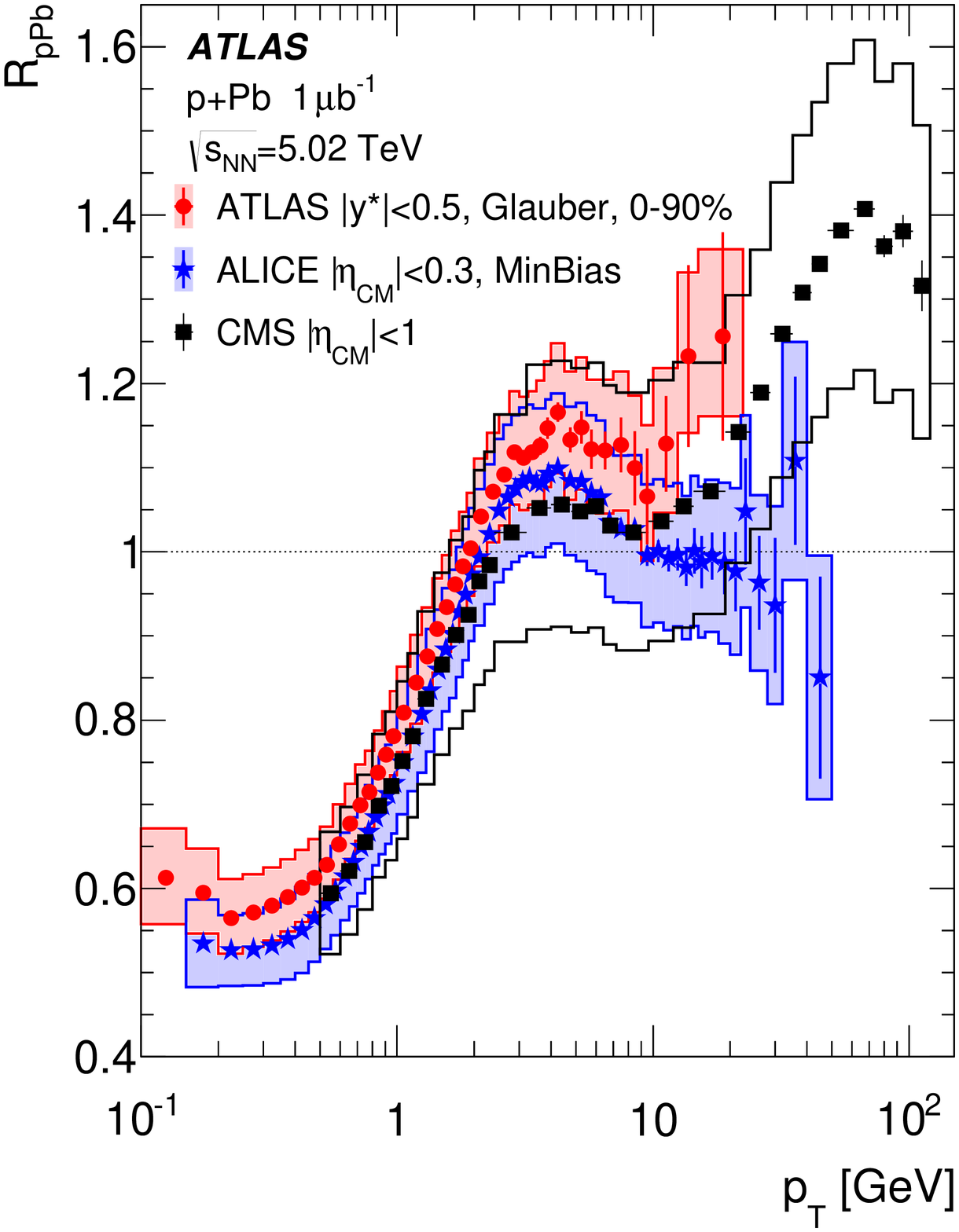}
\caption{\rpA values as a function of \pT\ in the 0--90\% centrality interval averaged over $|\ystar|<0.5$, are compared to the minimum-bias (0--100\%) results from a different pseudorapidity range in the centre-of-mass system: ALICE for $|\eta_{\mathrm{CM}}|<0.3$~\cite{ALICE:2012mj} and CMS for $|\eta_{\mathrm{CM}}|<1$~\cite{CMS-PAS-HIN-12-017}. The \avgTpb value for the ATLAS centrality correction is calculated with the \Gl model. The total systematic uncertainties, which include the uncertainty in \avgTpb,  are indicated by lines of the same colour. Strict quantitative agreement is not expected as each measurement uses different rapidity intervals for the centrality determination and apply different event selection criteria to reject diffractive collisions.} 
\label{fig:RpPb_vs_ALICE}
\end{center}
\end{figure}

The transverse momentum dependence of \rpA for the rapidity range $-1.8<\ystar<1.3$ and for the 0--90\% centrality interval is shown in Figure~\ref{fig:RpPb_0_90_G} 
for the \Gl and \GG calculations of \avgTpb. The 0--90\% \avgTpb values which are given in Table~\ref{tbl:NucTPb} are similar for all three estimations, therefore the curves in all three panels show little difference. For $\pT>8$~\GeV, \rpPb is consistent with unity for all three models in the range of statistical and systematic uncertainties. The \rpPb values obtained using the \Gl\ model for the \avgTpb\ calculation is compared to the ALICE~\cite{ALICE:2012mj} and CMS~\cite{CMS-PAS-HIN-12-017} measurements in Figure~\ref{fig:RpPb_vs_ALICE}. 
The results show the same basic features for the nuclear modification factors, although strict quantitative agreement is not expected as each measurement uses different rapidity intervals for the centrality determination and apply different event selection criteria to reject diffractive collisions.

\begin{figure}[t!]
\begin{center}
\includegraphics[width=0.98\textwidth]{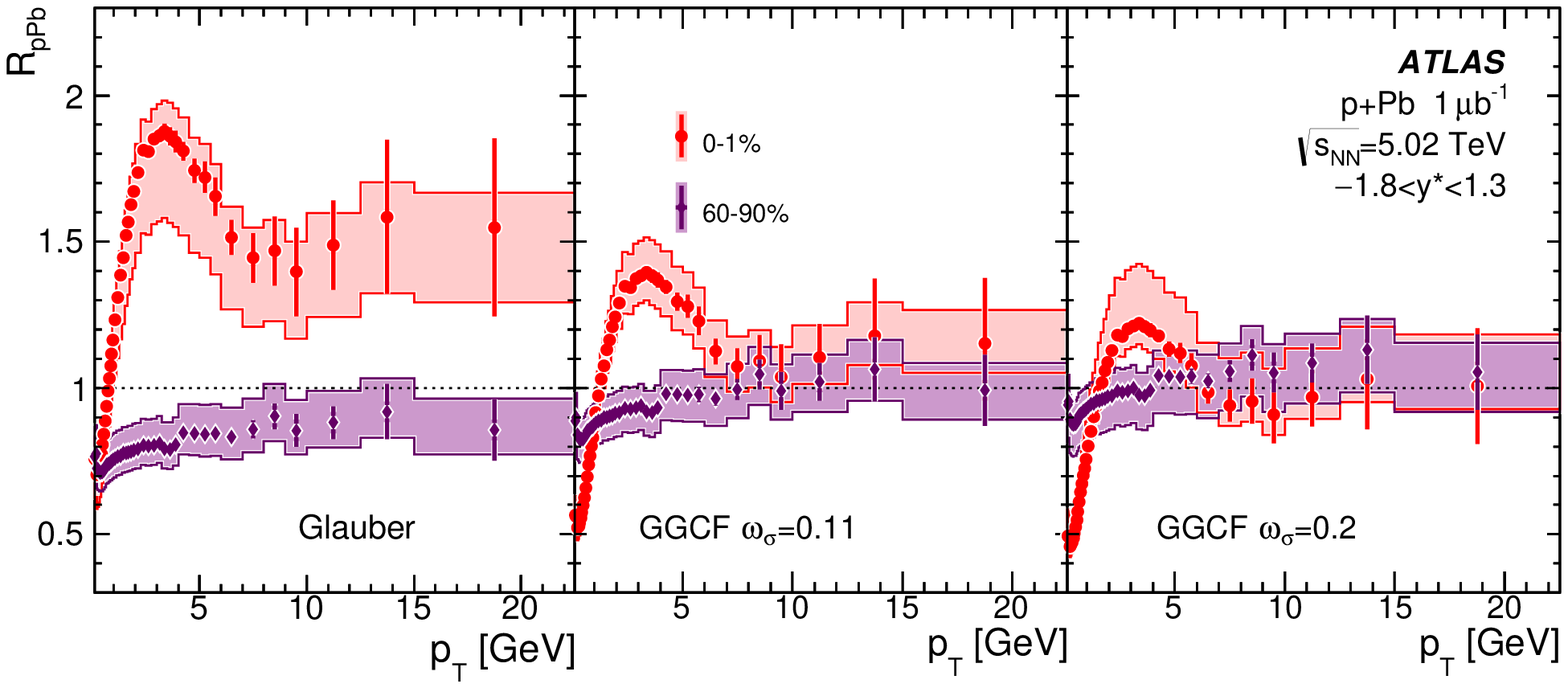}    
\includegraphics[width=0.98\textwidth]{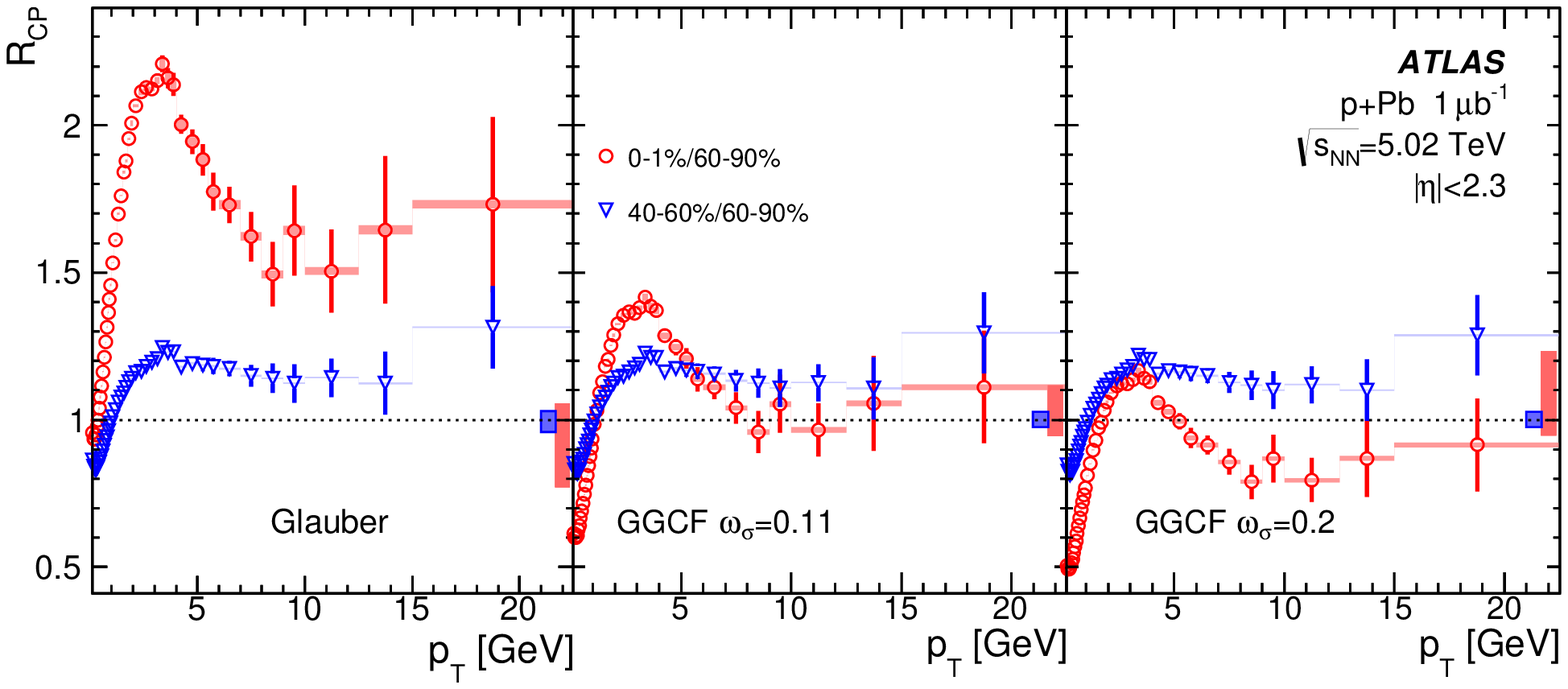}       
\caption{(top row) \rpA\ as a function of \pT extracted from the invariant yields integrated over $-1.8<\ystar<1.3$ 
for the 0--1\% and 60--90\% centrality intervals, and for different geometrical models used to calculate \avgTpb: \Gl, \GG\ \ggcfA\ and \GG\ \ggcfB;
(bottom row) \rcp for 0--1\% and 40--60\% central collisions with respect to the 60--90\% centrality interval, also for the geometrical models, which are used to calculate \avgTpb.
Statistical errors are indicated with vertical lines and the systematic uncertainties in the invariant yields are indicated by a shaded area. The total systematic uncertainty, which includes the uncertainty in \avgTpb are indicated by lines of the same colour. The systematic uncertainty in the ratios of \avgTpb are indicated by boxes of the same colour.} 
\label{fig:RpPb_2CBins}
\end{center}
\end{figure}    

The \rpPb and \rcp values are shown in Figure~\ref{fig:RpPb_2CBins} as a function of the charged-particle \pT in different centrality intervals and for different geometrical models used to calculate the value of \avgTpb. The data are integrated over $-1.8<\ystar<1.3$ for \rpPb and $|\eta|<2.3$ for \rcp. The data from the 0--1\% centrality interval show similar features in all panels. 
Both \rpA and \rcp increase with transverse momentum, reaching a maximum value at approximately $\pT\sim3$~\GeV\ and then decrease until reaching $\pT\sim8$~\GeV. 
Above this value, the ratios are approximately constant within the experimental uncertainties. The \rpPb and \rcp distributions in the region of the peak, $1<\pT<8$~\GeV, have larger  values for central events than for peripheral events. The magnitude of the peak depends quantitatively on the choice of geometrical model: the results obtained using the \Gl\ model have larger peak values than either of the \GG\ models. The magnitude of the peak relative to the constant (plateau) region ($\pT\gtrsim8$~\GeV) is compatible for \rcp and \rpPb given the systematic uncertainties. The peripheral events show a smaller rise at low \pT. There is also only a slight indication of a peak at $\pT\sim3$~\GeV\ in \rcp\ and no pronounced indication of a peak in the \rpPb. The magnitude of \rpA and \rcp in the constant region ($\pT\gtrsim8$~\GeV) is significantly above unity in the most central collisions for the Glauber model. In contrast, plateau regions are consistent with unity for \GG\ with \ggcfA\ and for \GG\ with \ggcfB. For the peripheral centrality interval, the plateau region is consistent with unity for \rpA\ and deviates from unity for \rcp. In peripheral collisions, \rpA and \rcp depend only weakly on the choice of \Gl\ or \GG\ model in all panels.

\newpage

\begin{figure}[b!]
\begin{center}      
\includegraphics[width=\textwidth]{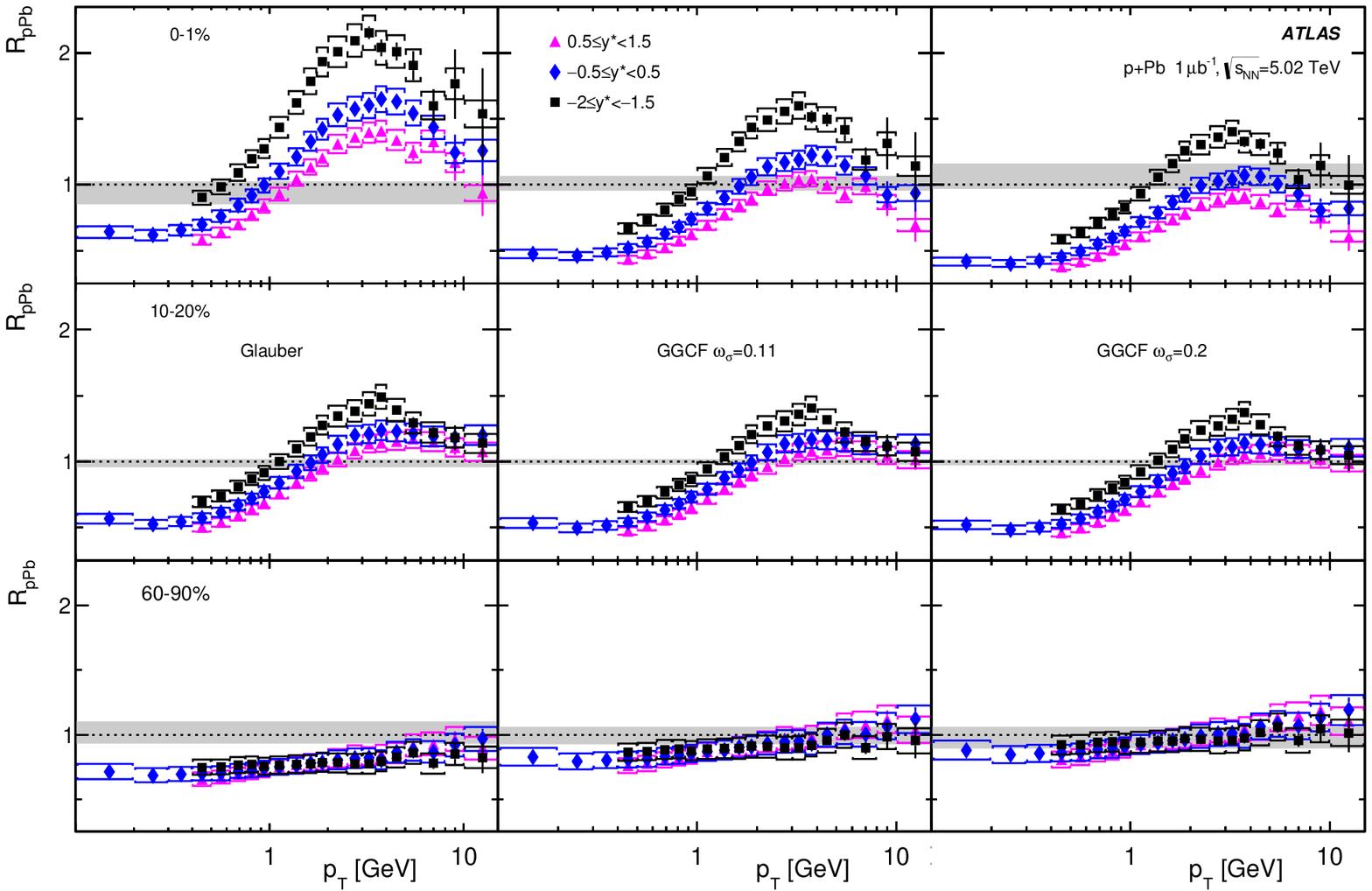}   
\caption{The \rpPb values for the 0--1\% (top panels), 10--20\% (middle panels) and 60--90\% (lower panels)centrality intervals. The data points are shown for three different rapidity intervals indicated in the legends. The columns correspond to the \Gl model (left), \GG model with \ggcfA\ (middle), and \GG\ model with \ggcfB\ (right). The grey band in each panel reflects the systematic uncertainty associated with the centrality interval and with the model assumption. Statistical uncertainties are shown with vertical bars and systematic uncertainties with brackets.
} 
\label{fig:rcp_EtaBins_CBins}
\end{center}
\end{figure}

\begin{figure}[h]
\begin{center}      
\includegraphics[width=\textwidth]{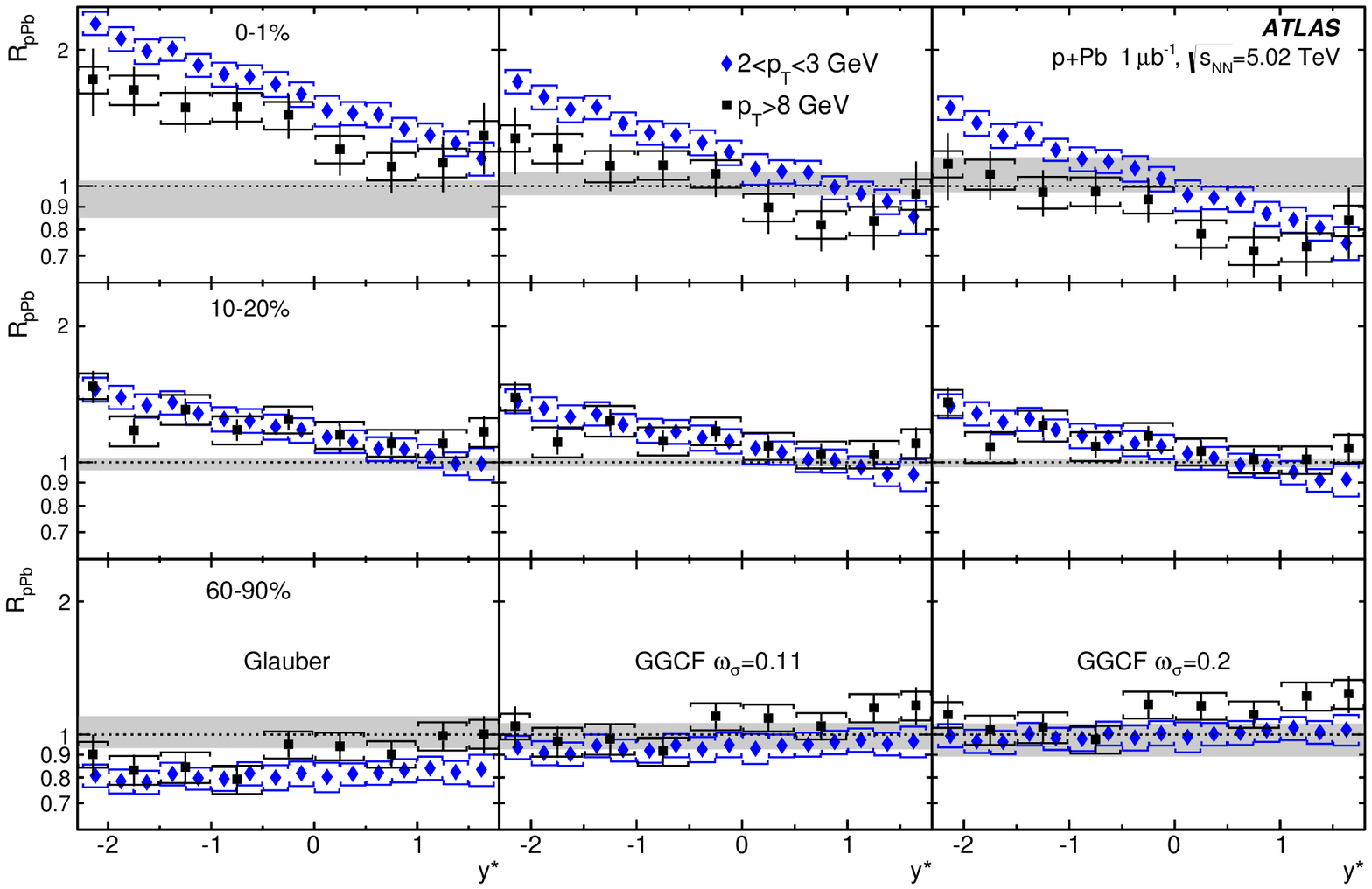}
\caption{The \rpPb values for the 0--1\% (top panels), 10--20\% (middle panels) and 60--90\% (lower panels) centrality intervals. The data points are shown for two different transverse momentum intervals indicated in the legends. The columns correspond to the \Gl\ model (left), \GG\ model with \ggcfA\ (middle), and \GG\ model with \ggcfB\ (right). The grey box in each panel reflects the systematic uncertainty associated with the centrality interval and with the model assumption. Statistical uncertainties are shown with vertical bars and systematic uncertainties with brackets.
} 
\label{fig:rpPb_ptBins_CBins}
\end{center}
\end{figure}

Figures~\ref{fig:rcp_EtaBins_CBins} 
and~\ref{fig:rpPb_ptBins_CBins} show \rpPb as a function of \pT and \ystar respectively. 
The three panels in each column correspond to the most central (upper panels), 
mid-central (middle panels) and most peripheral (lower panels) centrality intervals. 
The three columns show the results from different geometrical models: 
\Gl\ (left), \GG\ with \ggcfA\ (middle), and \GG\ with \ggcfB\ (right). 
The grey box on each axis reflects the fractional systematic uncertainty corresponding to the centrality interval and
geometric model, which applies to all data points in the
panel. The systematic uncertainties in the invariant yields
are indicated with boxes, and the vertical bars reflect the statistical
uncertainty at each point.
Figure~\ref{fig:rcp_EtaBins_CBins} shows \rpPb as a function of \pT. In the peripheral collisions, \rpPb is close to unity and shows almost no \ystar dependence. The \rpPb values in the 10--20\% and 0--1\% centrality classes exhibit a  stronger \ystar dependence. To illustrate the \ystar dependence, Figure~\ref{fig:rpPb_ptBins_CBins} shows the value of \rpPb\ measured for $2<\pT<3$~\GeV\ (peaking region) compared to the value measured for $\pT>8$~\GeV\ (the plateau region) as a function of \ystar, for different centrality intervals and geometrical models. In both regions, \rpPb increases with \ystar towards the Pb-going direction and with increasingly central collisions. The variation of \rpPb with centrality is much larger for the peaking region than for the plateau region. The \rpPb values in the two centrality intervals have similar variations as a function of \ystar.

\section{Conclusions} 
\label{sec:conclusions}

This \papertype\ presents measurements of the per-event charged-particle multiplicity in $1$~$\mu {\textrm b}^{-1}$ of \pPb collisions at $\sqn=5.02$\,\TeV\ with the ATLAS detector at the LHC. The differential particle yields in \pPb collisions are compared with those in \pp collisions using the nuclear modification factor, \rpPb. The \pp reference cross sections at $\sqn=5.02$\,\TeV\ are constructed by interpolation of measurements performed
at $\sqs=2.76$~\TeV\ and 7~\TeV.  The measurements of \rpPb are presented in the centre-of-mass frame  in the rapidity range $-2.3<\ystar<1.8$ and transverse momentum $0.1<\pT<22$\,\GeV.  The measurements of $\rcp$ are presented in the laboratory frame 
over the pseudorapidity range $-2.3<\eta<2.3$ and the same transverse momentum region. 
The results for \rpPb and \rcp are presented as a function of transverse momentum and centrality in different \ystar and $\eta$ intervals and also as a function of rapidity for different \pT\ intervals. The results are using two choices of  geometric model (\Gl\ and Glauber--Gribov colour-fluctuation model with \ggcfA~and~\ggcfB) for the calculation of the nuclear thickness function \avgTpb in the selected centrality intervals.

The measured nuclear modification factors are observed to increase with transverse momentum from $0.1$~\GeV\ to a peak value at $\pT\sim3$~\GeV, at which point they decrease slowly up to $\pT\sim8$~\GeV. Above $\pT\sim8$~\GeV\ the nuclear modification factors are constant within the experimental uncertainties.

The magnitude of the peak strongly depends both on rapidity and
centrality. It increases from the proton beam direction to the Pb beam
direction and from peripheral to central collisions. The constant
region above $\pT\approx8$~\GeV\ is less sensitive to the different
centrality and (pseudo)rapidity intervals. Measurements of the
absolute magnitudes of \rpPb\ integrated over centrality and averaged
over rapidity are similar for different geometric models, although
their centrality dependence is strongly influenced by the choice of
geometric model. Such behaviour is directly related to the
multiplicity dependence of the particle production. In particular,
there is an enhancement of protons with respect to pions at
intermediate \pT, as observed by other experiments at the LHC as well as at lower energies.

The momentum and rapidity dependence of the nuclear modification factor measured in \pPb collisions assist in determining the correct theoretical description of the cold nuclear matter effects. The results will also be important for constraining the choice of \Gl or \GG model parameters suitable to use in determining the average values for the number of participating nucleons and the nuclear thickness function in \pPb collisions.

\section*{Acknowledgements}


We thank CERN for the very successful operation of the LHC, as well as the
support staff from our institutions without whom ATLAS could not be
operated efficiently.

We acknowledge the support of ANPCyT, Argentina; YerPhI, Armenia; ARC, Australia; BMWFW and FWF, Austria; ANAS, Azerbaijan; SSTC, Belarus; CNPq and FAPESP, Brazil; NSERC, NRC and CFI, Canada; CERN; CONICYT, Chile; CAS, MOST and NSFC, China; COLCIENCIAS, Colombia; MSMT CR, MPO CR and VSC CR, Czech Republic; DNRF and DNSRC, Denmark; IN2P3-CNRS, CEA-DSM/IRFU, France; GNSF, Georgia; BMBF, HGF, and MPG, Germany; GSRT, Greece; RGC, Hong Kong SAR, China; ISF, I-CORE and Benoziyo Center, Israel; INFN, Italy; MEXT and JSPS, Japan; CNRST, Morocco; FOM and NWO, Netherlands; RCN, Norway; MNiSW and NCN, Poland; FCT, Portugal; MNE/IFA, Romania; MES of Russia and NRC KI, Russian Federation; JINR; MESTD, Serbia; MSSR, Slovakia; ARRS and MIZ\v{S}, Slovenia; DST/NRF, South Africa; MINECO, Spain; SRC and Wallenberg Foundation, Sweden; SERI, SNSF and Cantons of Bern and Geneva, Switzerland; MOST, Taiwan; TAEK, Turkey; STFC, United Kingdom; DOE and NSF, United States of America. In addition, individual groups and members have received support from BCKDF, the Canada Council, CANARIE, CRC, Compute Canada, FQRNT, and the Ontario Innovation Trust, Canada; EPLANET, ERC, FP7, Horizon 2020 and Marie Sk{\l}odowska-Curie Actions, European Union; Investissements d'Avenir Labex and Idex, ANR, R{\'e}gion Auvergne and Fondation Partager le Savoir, France; DFG and AvH Foundation, Germany; Herakleitos, Thales and Aristeia programmes co-financed by EU-ESF and the Greek NSRF; BSF, GIF and Minerva, Israel; BRF, Norway; Generalitat de Catalunya, Generalitat Valenciana, Spain; the Royal Society and Leverhulme Trust, United Kingdom.

The crucial computing support from all WLCG partners is acknowledged gratefully, in particular from CERN, the ATLAS Tier-1 facilities at TRIUMF (Canada), NDGF (Denmark, Norway, Sweden), CC-IN2P3 (France), KIT/GridKA (Germany), INFN-CNAF (Italy), NL-T1 (Netherlands), PIC (Spain), ASGC (Taiwan), RAL (UK) and BNL (USA), the Tier-2 facilities worldwide and large non-WLCG resource providers. Major contributors of computing resources are listed in Ref.~\cite{ATL-GEN-PUB-2016-002}.

\printbibliography

\newpage 
\begin{flushleft}
{\Large The ATLAS Collaboration}

\bigskip

G.~Aad$^\textrm{\scriptsize 87}$,
B.~Abbott$^\textrm{\scriptsize 114}$,
J.~Abdallah$^\textrm{\scriptsize 65}$,
O.~Abdinov$^\textrm{\scriptsize 12}$,
B.~Abeloos$^\textrm{\scriptsize 118}$,
R.~Aben$^\textrm{\scriptsize 108}$,
O.S.~AbouZeid$^\textrm{\scriptsize 138}$,
N.L.~Abraham$^\textrm{\scriptsize 150}$,
H.~Abramowicz$^\textrm{\scriptsize 154}$,
H.~Abreu$^\textrm{\scriptsize 153}$
% ...
\bigskip
\\
% ...
$^{12}$ Institute of Physics, Azerbaijan Academy of Sciences, Baku, Azerbaijan\\
% ...
$^{65}$ University of Iowa, Iowa City IA, United States of America\\
% ...
$^{87}$ CPPM, Aix-Marseille Universit{\'e} and CNRS/IN2P3, Marseille, France\\
% ...
$^{108}$ Nikhef National Institute for Subatomic Physics and University of Amsterdam, Amsterdam, Netherlands\\
% ...
$^{114}$ Homer L. Dodge Department of Physics and Astronomy, University of Oklahoma, Norman OK, United States of America\\
% ...
$^{118}$ LAL, Univ. Paris-Sud, CNRS/IN2P3, Universit{\'e} Paris-Saclay, Orsay, France\\
% ...
$^{138}$ Santa Cruz Institute for Particle Physics, University of California Santa Cruz, Santa Cruz CA, United States of America\\
% ...
$^{150}$ Department of Physics and Astronomy, University of Sussex, Brighton, United Kingdom\\
% ...
$^{153}$ Department of Physics, Technion: Israel Institute of Technology, Haifa, Israel\\
$^{154}$ Raymond and Beverly Sackler School of Physics and Astronomy, Tel Aviv University, Tel Aviv, Israel\\
% ...
\end{flushleft}

\end{document}